\documentclass[12pt,preprint]{aastex}

\usepackage{natbib}

\shorttitle{Near-Infrared Survey of EGOs}

\shortauthors{H.-T. Lee et al.}

\begin{document}

%\title{Investigating Outflows of Extended Green Objects}
\title{Near-Infrared H$_{2}$ and Continuum Survey of Extended Green Objects}

\author{Hsu-Tai Lee\altaffilmark{1} \email{htlee@asiaa.sinica.edu.tw}}

\author{Michihiro Takami\altaffilmark{1}}

\author{Hao-Yuan Duan\altaffilmark{2,1}}

\author{Jennifer Karr\altaffilmark{1}}

\author{Yu-Nung Su\altaffilmark{1}}

\author{Sheng-Yuan Liu\altaffilmark{1}}

\author{Dirk Froebrich\altaffilmark{3}}

\author{Cosmos C. Yeh\altaffilmark{4,1}}

\altaffiltext{1}{Institute of Astronomy and Astrophysics, Academia Sinica, P.O. Box 23-141, Taipei 106, Taiwan}

\altaffiltext{2}{Department of Earth Sciences, National Taiwan Normal University, Taiwan}

\altaffiltext{3}{Centre for Astrophysics and Planetary Science, University of Kent, Canterbury, CT2 7NH, U.K.}

\altaffiltext{4}{Institute of Astronomy and Department of Physics, National Tsing Hua University, Hsinchu, Taiwan}

\begin{abstract}

The $Spitzer$ GLIMPSE survey has revealed a number of ``Extended Green Objects" (EGOs) which display extended emission at 4.5~$\micron$.  These EGOs are potential candidates for high mass protostellar outflows.  We have used high resolution ($< 1\arcsec$) H$_{2}$ 1-0 S(1) line, $K$, and $H$-band images from the United Kingdom Infrared Telescope to study 34 EGOs to investigate their nature.  We found that 12 EGOs exhibit H$_{2}$ outflows (two with chains of H$_{2}$ knotty structures; five with extended H$_{2}$ bipolar structures; three with extended H$_{2}$ lobes; two with pairs of H$_{2}$ knots).  In the 12 EGOs with H$_{2}$ outflows, three of them exhibit similar morphologies between the 4.5~$\micron$ and H$_{2}$ emission.  However, the remaining 9 EGOs show that the H$_{2}$ features are more extended than the continuum features, and the H$_{2}$ emission is seldom associated with continuum emission.  Furthermore, the morphologies of the near-infrared continuum and 4.5~$\micron$ emission are similar to each other for those EGOs with $K$-band emission, implying that at least a part of the IRAC-band continuum emission of EGOs comes from scattered light from the embedded YSOs.

\end{abstract}


\section{INTRODUCTION}%1

Massive stars play an important role in the evolution of galaxies.  Once they are born, their outflows, intense UV radiation, and stellar winds influence the nearby interstellar medium, and their feedback could induce the next generation of star formation \citep{zin07,lee05,lee07,kar09}.  However, the formation mechanism of massive stars is still unclear.  One of reasons for this is that massive protostars are very embedded, and hence can only be seen in infrared and radio.  Since the $Spitzer$ space telescope was launched, it has provided a good chance to study this early phase of star formation.  $Spitzer$ provides better resolution than previous infrared space telescopes ($\sim2\arcsec$ at IRAC bands), and is able to survey large regions.  In the Galactic Legacy Infrared Mid-Plane Survey Extraordinaire (GLIMPSE) I survey at IRAC wavelengths \citep[$10\arcdeg < |l| < 65\arcdeg$; $|b| < 1\arcdeg$,][]{ben03}, \citet{cyg08} identified more than 300 massive outflow candidates in the Galactic Plane.  These candidates show excess emission at 4.5~$\micron$, and have been dubbed ``Extended Green Objects (EGOs)'' due to the common practice of assigning blue, green and red to 3.6, 4.5, and 8~$\micron$ bands in images.  The most common explanation is that they are outflows from massive young stars, due to their association with IRDCs, thought to be the birthplace of massive stars, and with Class II 6.7 GHz masers which are radiatively pumped by massive YSOs \citep{min03}.  The 4.5~$\micron$ emission may come from H$_{2}$ ($\upsilon$ = 0--0, S(9, 10, 11)) and/or CO ($\upsilon$ = 1--0) band heads \citep{rea06} which can be excited by shocks from outflows \citep{cyg08}.

Since their identification, there have been several attempts to investigate the nature of these EGOs.  Approximately two-third of EGOs are associated with Class I 44~GHz methanol masers \citep{che09} which are likely to be an outflow tracer.  Later, high resolution VLA observations showed that Class II methanol masers are tightly concentrated around 24~$\micron$ sources, and Class I masers tracing molecular outflows are widely distributed around the diffuse 4.5~$\micron$ emission \citep{cyg09}.  \citet{che10} and \citet{cyg09} used outflow tracers (HCO$^{+}$) to survey EGOs, and their line profiles also indicate the presence of outflows.  EGOs \citep[or green fuzzies,][]{cha09} have been seen in known massive star forming regions \citep[e.g.,][]{dav07,smi06,she07,cyg07,var11}, IRDC cores \citep{cha09}, and jets associated with low-mass protostars \citep[e.g.,][]{zha09,tak10}.

Molecular line interferometric observations can also help to identify outflows from massive YSOs, and provide higher resolution and kinematic information for outflows \citep[see review by][]{beu05}.  EGO~G11.92$-$0.61 and G19.01$-$0.03 were observed using SMA and CARMA \citep{cyg11}.  Both of them have bipolar molecular outflows, coincident with the 4.5~$\micron$ lobes.

The above mentioned studies give evidence that some EGOs are related to outflows; however it is unclear if the 4.5~$\micron$ emission in EGOs comes from line emission or not.  The best way to answer this question is to take spectra of EGOs in the IRAC 4.5~$\micron$ band.  The Infrared Spectrograph (IRS) on-board $Spitzer$ can observe only wavelengths longer than 5.3~$\micron$, so IRS itself can not directly determine the source of diffuse 4.5~$\micron$ emission.  To determine the origin of what they denote as ``green fuzzies'', \citet{deb10} took spectra of two EGOs (EGO~G19.88$-$0.53 and EGO~G49.27$-$0.34) in $L$- (2.95 -- 4.12~$\micron$) and $M$- (4.47 -- 5.75~$\micron$) bands.  One of them (EGO G19.88-0.53) does exhibit pure molecular hydrogen line emission at 4.695~$\micron$ ($\upsilon = 0-0$ $S(9)$), which dominates in the wavelength range 4--5~$\micron$ (no CO band heads are detected).  This result supports the idea that the green fuzzy emission is excited by outflow shocks.  However, the other source (EGO G49.27-0.34) only shows continuum emission in the above spectral ranges without line emission.  They suggested that the green fuzzy emission is the result of extinction and a magnification in the color stretch in the 4.5~$\micron$ band.%  In addition to the spectra of green fuzzy emission, \citet{deb10} also took spectra of the two point sources in $L$ and $M$ bands.  Both of them show XCN, CO, and OCS absorption features at 4.62, 4.67, and 4.9~$\micron$, respectively, and these features indicate they are likely embedded MYSOs \citep{gib04}.

In this paper, we use H$_{2}$, $H$-, and $K$-band observations from the United Kingdom Infrared Telescope (UKIRT) to study the near-infrared properties of EGOs.  Near-infrared H$_{2}$ 1-0 S(1) observations have been used to study shock-excited outflows from YSOs \citep{dav09,ste09}, and provide a good resolution ($\sim$1$\arcsec$) which allows us to study the morphologies of H$_{2}$ outflows in detail.  Compared with continuum images of Spitzer, our continuum-subtracted H$_{2}$ images, which focus on shock-excited H$_{2}$ line emission, are helpful for determining if the EGOs are associated with H$_{2}$ outflows or not. % Regarding statistic result,e.g., what the fraction of EGO with H$_{2}$ outflow is, we will discuss in the next paper.

%H$_{2}$ observations can be used to indicate that shocked regions or fluorescently excited gas (T=$\sim$2000~K, n$_{H2}$ $>$ 10$^{3}$ cm$^{-3}$) in star-forming regions.  Thus, it is helpful to trace outflows from embedded YSOs.  In this paper, we use continuum-subtracted H$_{2}$ images to study outflow features.  Compared with low resolution and continuum images of $Spitzer$, our continuum-subtracted H$_{2}$ images, which dedicate to shock-excited H$_{2}$ line emission, can tell if EGOs are outflows from YSOs or not.  Near-infrared H$_{2}$ observations have been used to study shock-excited outflows from YSOs \citep{dav10,dav09,dav04,var10,var11,kum02}, and provide better resolution ($\sim1\arcsec$) which allow us to study the morphologies of H$_{2}$ outflows in detail.  %However, the H$_{2}$ observations may suffer higher extinction, compared with the $Spitzer$ mid-infrared observations.  

This paper is organized as follows.  We describe the observations and results in Section 2 and 3, respectively.  We discuss the near-infrared properties of EGOs in Section 4, and summarize our conclusions in Section 5.

\section{TARGET SELECTION, OBSERVATIONS, AND DATA REDUCTION}%2 

Narrow-band images of H$_{2}$ $\upsilon$ =1-0 S(1) emission (2.122~$\micron$) were obtained through the UKIRT Widefield Infrared Survey for H$_{2}$ \citep[UWISH2,][]{fro11a}.  UWISH2 began in July 2009 and was completed in August 2011; it surveyed the inner Galactic Plane ($10\arcdeg \lesssim l \lesssim 65\arcdeg; -1.3\arcdeg \lesssim b \lesssim +1.3\arcdeg$) with the Wide Field Camera (WFCAM) at UKIRT.  The observations were taken with a narrow band filter ($\Delta\lambda=0.021~\micron$) at 2.122~$\micron$ with 720 seconds integration time per pixel.  The median seeing in H$_{2}$ images is $\sim$0.7$\arcsec$; only a few images were taken with seeing $> 1\arcsec$.  The 5$\sigma$ detection limit of point sources is $\sim$18 mag in $K$, and the surface brightness limit is $10^{-19}$~W~m$^{-2}$~arcsec$^{-2}$ at the typical seeing \citep{fro11a}.

The $K$- and $H$-band images were obtained through the WSA archive for the UKIRT Infrared Deep Sky Survey Galactic Plane Survey \citep[UKIDSS GPS,][]{luc08}.  GPS is one of the five surveys within UKIDSS, and will survey more than 1800 square degrees of the northern equatorial Galactic plane in $J$, $H$ and $K$-bands with WFCAM.  The 5 $\sigma$ detection limit of point sources are $H$=19 and $K$=18.5~mag, and a seeing $<$ 1$\arcsec$ in $K$ band is requested by the survey.  $K$-band (2.2~$\micron$) images were used to construct our continuum-subtracted H$_{2}$ images of EGOs, and we compared their morphologies with the H$_{2}$ and continuum ($H$-band and 4.5~$\micron$) emission.

%Target Selection & Observations

%The data sets for 34 EGOs were obtained from the above survey.  These were selected from the MYSO outflow candidates in Tables 1-4 of \citet{cyg08}.  Tables 1 and 2 of \citet{cyg08} include sources of ``likely'' MYSO outflow candidates, and Tables 3 and 4 of \citet{cyg08} include ``possible'' MYSO outflow candidates.  The difference between ``likely'' and ``possible'' MYSO outflow candidates is based primarily on the angular extent of the extended 4.5~$\micron$ emission in three-color images.  If a source may be caused by an image artifact from a bright IRAC source that could be confused with real extended 4.5~$\micron$ emission, then it is considered a ``possible'' MYSO outflow candidate.  Tables 1 and 3 in \citet{cyg08} list integrated IRAC 4 bands and MIPS 24~$\micron$ flux densities for EGOs in all five bands.  However, Tables 2 and 4 in \citet{cyg08} only list integrated 4.5~$\micron$ flux densities for EGOs, because these sources may suffer confusion and/or saturation in some bands.

The UWISH2 data for 12 EGOs were obtained in the earliest phase of the survey as high priority targets.  These targets were selected as follows.  We investigated GPS $K$-band images of EGOs in Table 1 of \citet{cyg08}, before the H$_{2}$ observations.  If there was $K$-band continuum emission within a 20$\arcsec$ radius of the EGO's positions, we included them as high priority targets for UWISH2 observations in 2009.  The criterion is not only restricted to the EGO's positions, but also extended to a 20$\arcsec$ radius, as the morphologies of H$_{2}$ outflows may be more extended than those of $K$-band continuum.  Besides these preselected 12 EGOs, another 22 EGOs were covered by the UWISH2 survey.  In this paper, we studied a total of 34 EGOs which were observed by UWISH2 in 2009.  Table~\ref{tab:h2} shows the list of these targets.  The GPS covers all the targets in $K$-band and $H$-band except for EGO G35.20-0.74 ($K$-band only).

%For these 22 EGOs, 7, 4, 2, and 9 sources belong to Tables 1, 2, 3, and 4 of \citet{cyg08}, respectively.
%  Fourteen EGOs showing extended emission in $K$-band were selected as high priority targets.  In the 2009 observations, 12 of 14 the high priority targets were observed.
% 1-21 and 1-34 as high priority targets are not observed in 2009

%Data reduction

Data reduction was done by the Cambridge Astronomical Survey Unit (CASU).  At CASU, the WFCAM pipeline is responsible for flat fielding and sky correction of the image data, and stacking frames as described by \citet{dye06}.  Reduced $K$-band and H$_{2}$ images were accessed via the Wide Field Astronomy Unit\footnote{http://www.roe.ac.uk/ifa/wfau/}.  

The continuum-subtracted H$_{2}$ images were obtained as follows.  Using IDL, we shifted and rotated each $K$-band image to match the coordinates of the corresponding H$_{2}$ image.  Then we compared the full width at half maximum (FWHM) of the $K$-band and H$_{2}$ images for each EGO, and smoothed the smaller FWHM to the larger one.  Before subtraction, we scaled the $K$-band fluxes to match their stellar fluxes to those in the H$_{2}$ image.  Then, a continuum-subtracted H$_{2}$ image of each EGO was generated by subtracting the $K$-band from the H$_{2}$ image.

%Spitzer observations

In addition to the H$_{2}$ and $K$-band images described above, $Spitzer$ IRAC and MIPS 24~$\micron$ images from the GLIMPSE \citep{chu09} and MIPSGAL \citep{car09} legacy surveys, respectively, were obtained from the archive data for comparison.  The data had been reduced with post Basic Calibrated Data (post-BCD) piplines developed by the Infrared Processing and Analysis Center (IPAC).  The IRAC images are used to compare the spatial distribution of the EGOs with continuum-subtracted H$_{2}$ and $K$-band images.  The MIPS 24~$\micron$ images will be used to identify YSOs from thermal dust emission.

\section{RESULTS}%3

%Of our 34 EGO sample, 22 EGOs are selected to discuss in the Appendix, because these EGOs show clear detections of positive (H$_{2}$ emission) or/and negative (continuum emission) valued features in the continuum-subtracted H$_{2}$ images (Figure~\ref{fig:G10.34-0.14} to \ref{fig:G58.09-0.34}).  We present the results for these 22 EGOs in this section.  These EGOs are shown together with the images of the IRAC bands (3.6, 4.5, and 8.0~$\micron$), MIPS 24~$\micron$, and $K$-band images.%  Detailed descriptions for individual objects are located in the Appendix.

Of our 34 object sample, 23 sources show clear detections of positive (i.e., H$_{2}$ emission) or/and negative (presumably due to continuum, as described below in detail) valued features, in the continuum-subtracted H$_{2}$ images.  Figures~\ref{fig:G10.34-0.14} - \ref{fig:G58.09-0.34} show their images in the IRAC bands (3.6, 4.5, and 8.0~$\micron$), MIPS 24~$\micron$, continuum-subtracted H$_{2}$, and $K$-band.  %

Table~\ref{tab:h2} summarizes the detections of H$_{2}$, $K$, and $H$-band emission.  Based on the continuum-subtraced H$_{2}$ images, 34 EGOs are categorized as follows: those associated with H$_{2}$ outflows (Y, 12 objects); those associated with possible H$_{2}$ outflows, (Y?, 4 objects);  and finally, those with no H$_{2}$ detection (N, 18 objects).  In addition, we include distances of EGOs \citep{che10,cyg09} in Table~\ref{tab:h2}.  Molecular Hydrogen Emission-Line Objects \citep[MHOs\footnote{http://www.jach.hawaii.edu/UKIRT/MHCat/},][]{dav10} numbers are assigned for newly identified H$_{2}$ outflows in Table~\ref{tab:mho}.

In Sections 3.1 and 3.2, we describe the details of the results for the H$_{2}$ and continuum emission ($K$ and $H$-band), respectively.  The details for the 23 selected individual objects are presented in the Appendix.

%Thus, only the 22 EGOs are shown and discussed in the Appendix (Figures~\ref{fig:G10.34-0.14} to \ref{fig:G58.09-0.34}).  Figure of each EGO shows three-color IRAC, 24~$\micron$ MIPS, continuum-subtracted H$_{2}$, and $K$-band images of the 22 EGOs.  These images are used for identifying morphologies of EGOs, the positions of MYSOs, morphologies of H$_{2}$ outflows, and scattered continuum emission, respectively.

%In our 34 EGO sample, almost every EGO is associated with a 24~$\micron$ source.  These 24~$\micron$ sources are likely embedded YSOs.  For those EGOs with H$_{2}$ outflows, the positions of the 24~$\micron$ sources are helpful to locate their outflow bases.  

\subsection{H$_{2}$ Emission}%3.1

In the continuum-subtracted H$_{2}$ images, 16 objects show positive valued features due to H$_{2}$ emission (seen in black in Figures~\ref{fig:G10.34-0.14} - \ref{fig:G58.09-0.34}).  Seven EGOs are associated with an extended bipolar structure or a set of aligned knots in these images (EGO G19.88-0.53, G35.04-0.47, G35.13-0.74, G35.15+0.80, G35.20-0.74, G35.79-0.17, and G35.83-0.20).  Two EGOs exhibit two H$_{2}$ knots (EGO 19.01-0.03 and G54.45+1.01).  Three EGOs only show one lobe in H$_{2}$ emission (EGO G11.92-0.61, G16.61-0.24, and G35.68-0.18).  Four EGOs show possible H$_{2}$ outflows (EGO G12.02-0.21, G12.42+0.50, G12.91-0.03, and G16.59-0.05), exhibiting either extended H$_{2}$ emission or an isolated H$_{2}$ knot.  The extended emission in the continuum-subtracted H$_{2}$ image of EGO G12.42+0.50 could be due to residuals from the subtraction of bright continuum emission of the YSOs. %or variability of the YSO.

In most of the above objects, the H$_{2}$ emission appears to be associated with a monopolar or bipolar H$_{2}$ outflow driven by a single protostar.  The protostellar outflows have a bipolar geometry in most cases, and the absence of the counterpart in some of the outflows is attributed to obscuration of the counterflow by extinction \citep[see][for review]{arc07,bal07}

However, EGO G19.88-0.53 and G35.13-0.74 seem to be associated with more than one H$_{2}$ outflow.  Generally, the H$_{2}$ outflows of the EGOs usually show small opening angles (highly collimated outflows).  In contrast, the H$_{2}$ outflow of EGO G35.20-0.74 shows an hourglass shape with a large opening angle, $\sim40\arcdeg$.

In most of the above objects, the H$_{2}$ emission is more extended than the 4.5~$\micron$ emission, and faint or absent at the regions where the 4.5~$\micron$ emission is bright.  The 4.5~$\micron$ emission shows a counterpart to the H$_{2}$ features in the following three EGOs.  In G19.88-0.53, diffuse 4.5~$\micron$ emission is observed at the base of the western lobe of the H$_{2}$ outflow.  \citet{deb10} conducted ground-based spectroscopy of this emission component at 2.9-4.1~$\micron$ and 4.5-5.8~$\micron$, and their spectra indicate that the emission in the IRAC 4.5~$\micron$ should be dominated by H$_{2}$ emission.  In G35.68-0.18, the H$_{2}$ feature ``B'' indicated in Figure~\ref{fig:G35.68-0.18} shows a faint counterpart in the 4.5~$\micron$ emission.  In G35.83-0.20, three bright H$_{2}$ knots in Figure~\ref{fig:G35.83-0.20} are also seen in the 4.5~$\micron$ image.  It is notable, however, the GLIMPSE image also contains a dozen brighter features associated with the 4.5~$\micron$ emission.

%For those EGOs with H$_{2}$ outflows, their morphologies of H$_{2}$ and 4.5~$\micron$ emission are different.  Generally, the 4.5~$\micron$ peaks are not associated with the H$_{2}$ emission.  The H$_{2}$ emission is more extended than the 4.5~$\micron$ emission, and only shows where the 4.5~$\micron$ emission is week or not detected.  In contrast, the distributions of 4.5~$\micron$ emission and $K$-band are similar.

%Another possibility is due to variability of YSOs.  The $K$-band images were obtained a few years prior to the H$_{2}$ images \citep{fro11a}.  If the extended $K$-band emission becomes fainter when H$_{2}$ images are taken, then the continuum-subtracted H$_{2}$ images around the YSOs would show negative values.  

%\citet{cyg09} performed a Class I 44 GHz methanol (CH$_{3}$OH) maser survey of 19 EGOs using VLA, and 17 of them show positive detections.  Seven EGOs with 44 GHz methanol maser detections overlap our sample.  We found that only two of the seven EGOs are associated with H$_{2}$ outflows (EGO G11.92-0.61 and G19.01-0.03).  Figure~\ref{fig:masers} shows the distributions of the methanol masers from \citet{cyg09}.  The majority of the methanol masers around EGO G11.92-0.61 are clustered at the H$_{2}$ lobe; however they are not distributed at the the position of the brightest part of the H$_{2}$ emission.  In EGO G19.01-0.03, most of the methanol masers gather around the northern lobe of the H$_{2}$ outflow, but no methanol maser is associated with the southern one.

\begin{figure}%24
\includegraphics[angle=0,scale=0.8]{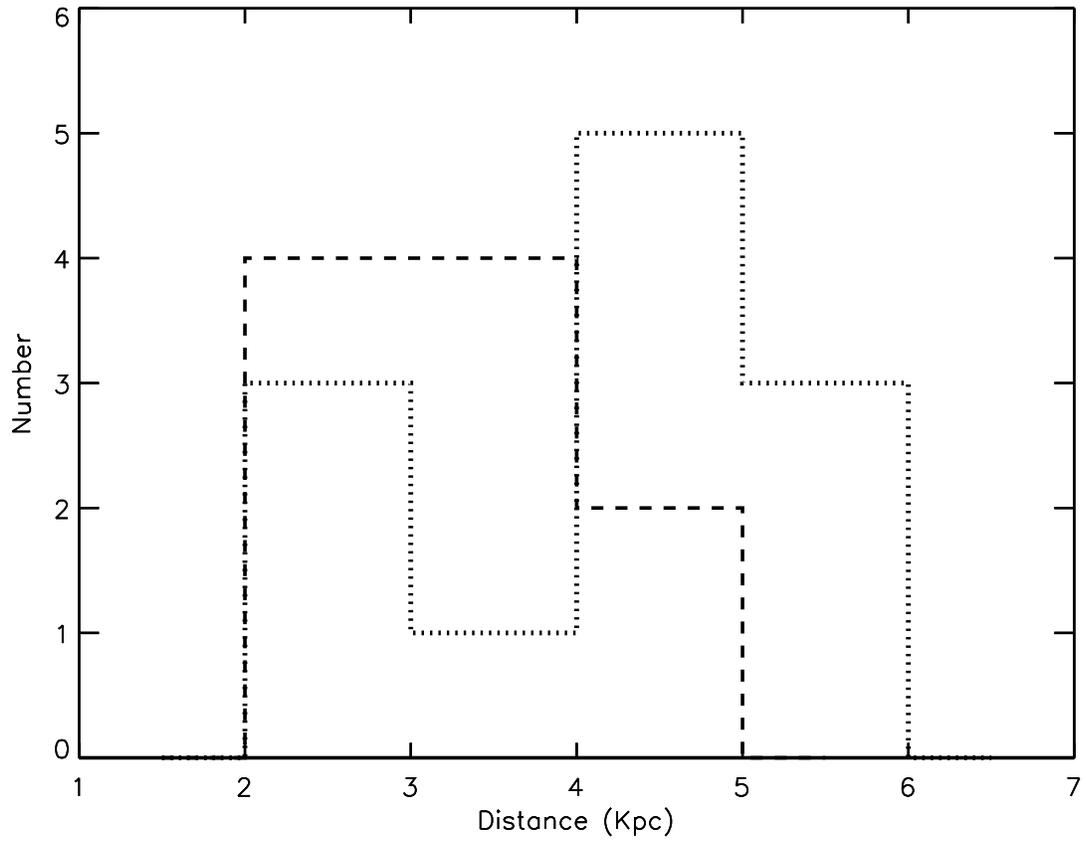}
\caption{Histograms of distances for EGOs with (dashed line) and without (dotted line) H$_{2}$ outflows.  Here, we only include those EGOs with distances in Table~\ref{tab:h2}, but exclude possible H$_{2}$ outflows in the histogram.}
\label{fig:hist}
\end{figure}

Figure~\ref{fig:hist} shows histograms of the distances for EGOs with H$_{2}$ (dashed line histogram) and without (dotted line histogram) detections.  We only include those EGOs with distances in Table~\ref{tab:h2}, but exclude possible H$_{2}$ outflows in the histogram.  In general, the distances of the EGOs without H$_{2}$ outflow detections seem to be larger than those with H$_{2}$ outflow detections.  Beyond four kpc, the numbers of H$_{2}$ outflow detections drop, but the numbers of non-detections increase.  Therefore, our H$_{2}$ outflow detection has a bias toward distance, and we probably miss H$_{2}$ outflow detections for some distant EGOs.

In the continuum-subtracted H$_{2}$ images (Figures~\ref{fig:G10.34-0.14} - \ref{fig:G58.09-0.34}), the positions of extended $K$-band emission (see Section 3.2 and Appendix for details) usually show negative valued features (seen in white in Figures~\ref{fig:G10.34-0.14} - \ref{fig:G58.09-0.34}) that are likely to represent continuum emission with a large infrared excess and high extinction, and may be light scattered from the YSOs.  This type of emission can also be seen around other H$_{2}$ outflow bases \citep{fro11b,var10}.  There are 12 EGOs associated with negative valued features around the positions of the EGOs in the continuum-subtracted H$_{2}$ images.  Six of them are associated with H$_{2}$ outflows (EGO G11.92-0.61, G16.61-0.24, G19.88-0.53, G35.13-0.74, G35.20-0.74, and G54.45+1.01), and the other six are not (EGO G10.34-0.14, G12.20-0.03, G19.61-0.12, G20.24+0.07, G28.83-0.25, and G58.09-0.34).%In the continuum-subtracted H$_{2}$ images, negative valued feature represents that the $K$-band continuum dominates the emission (seen in white in Figures~\ref{fig:G10.34-0.14} - \ref{fig:G58.09-0.34}).  

%Of the 23 EGOs with the $K$-band diffuse emission, 7 of them are also detected in $H$-band.  Figure~\ref{fig:hk} shows the 7 EGOs with the $H$-band emission and includes their images in $H$, $K$, 3.6~$\micron$, and 4.5~$\micron$.  For those EGOs with $H$-band detections, the diffuse emission becomes apparent toward longer wavelengths.  The emission around the point sources also become brighter as well, which are faint or not detected in near-infrared, and this is due to extinction and the SEDs of the embedded YSOs which are brighter toward longer wavelengths.  The near-infrared features are not as clear as those in mid-infrared.  However, the near- and mid-infrared morphologies are similar (excluding the mid-infrare emission around the point sources), if we only compare the near-infrared features with the bright mid-infrared ones.

\subsection{Continuum Emission}%3.2

Figures~\ref{fig:G10.34-0.14} - \ref{fig:G58.09-0.34} show that at least 9 of the 23 EGOs show similar morphologies in the $K$ and 4.5~$\micron$ emission (EGO G12.42+0.50, G19.01-0.03, G28.83-0.25, G35.13-0.74, G35.15+0.80, G35.20-0.74, G35.68-0.18, G35.83-0.20, and G54.45+1.01), although the signal to noise ratios of the $K$-band emission are lower than those of the 4.5~$\micron$ emission.  In the remaining 13 EGOs, 10 EGOs exhibit extended $K$-band emission at the position of the EGOs (EGO G10.34-0.14, G11.92-0.61, G12.20-0.03, G12.91-0.03, G16.61-0.24, G19.61-0.12, G19.88-0.53, G20.24+0.07, G35.79-0.17, and G58.09-0.34), and three EGOs show a $K$-band point source around the 4.5~$\micron$ peak of each EGO (EGO G12.02-0.21, G16.59-0.05, and G35.04-0.47).  One or more 24~$\micron$ sources are usually located near the 4.5~$\micron$ peak of the majority of the EGOs.

%All of the 22 EGOs are associated with one or more 24~$\micron$ sources, and the 24~$\micron$ sources are usually located at or around to the EGOs.  

%The $K$-band and 4.5~$\micron$ extended emission usually peaks around the 24~$\micron$ sources, and becomes fainter farther from the peaks.  

\begin{figure}%25
\includegraphics[angle=0,scale=0.59]{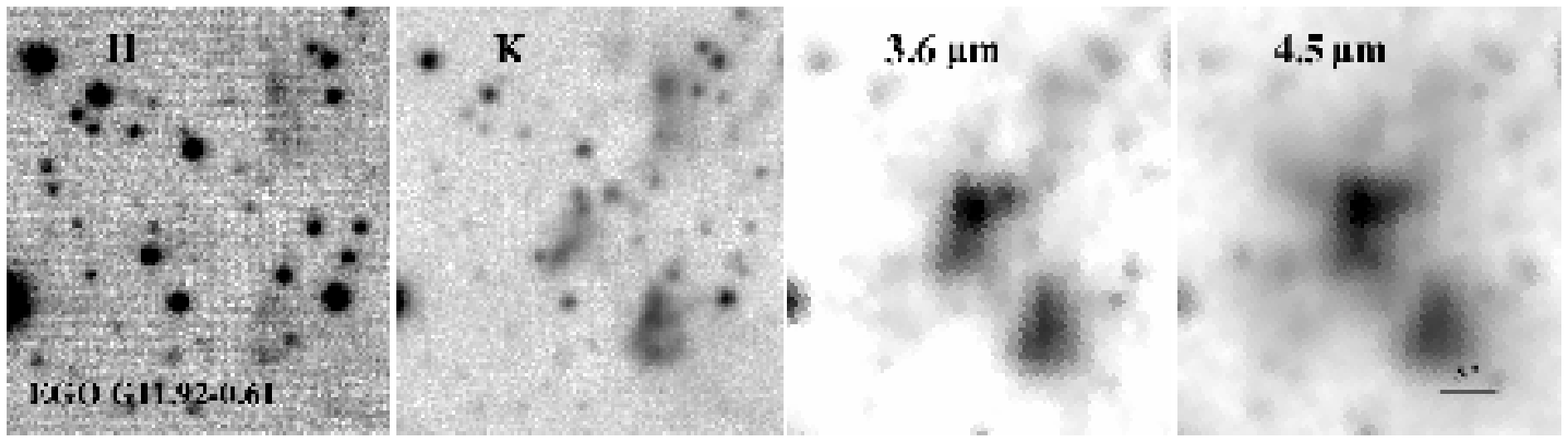}
\includegraphics[angle=0,scale=0.59]{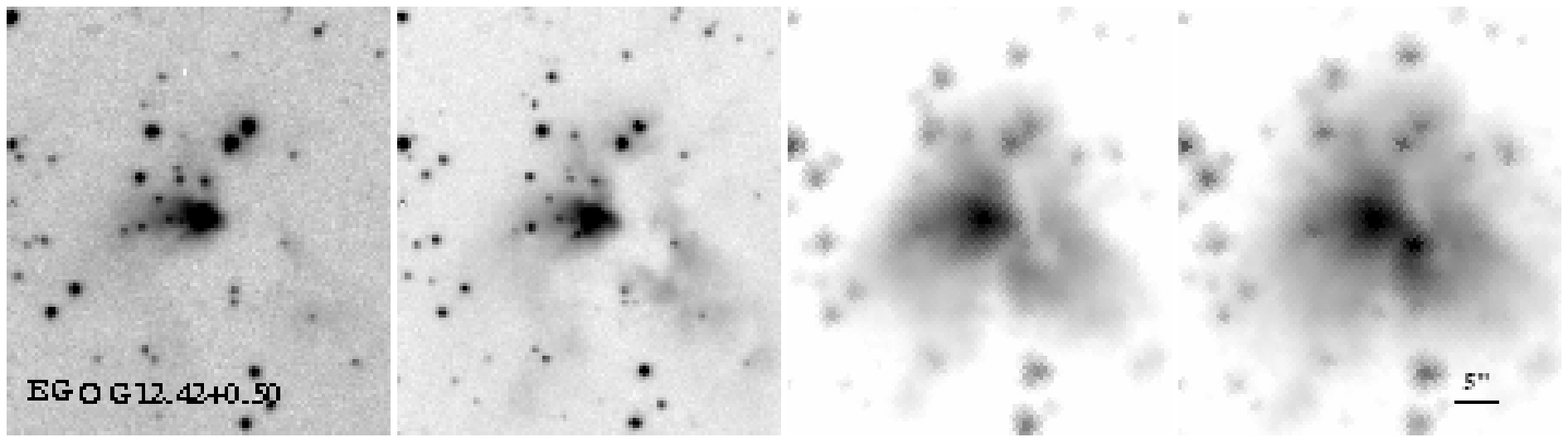}
\includegraphics[angle=0,scale=0.59]{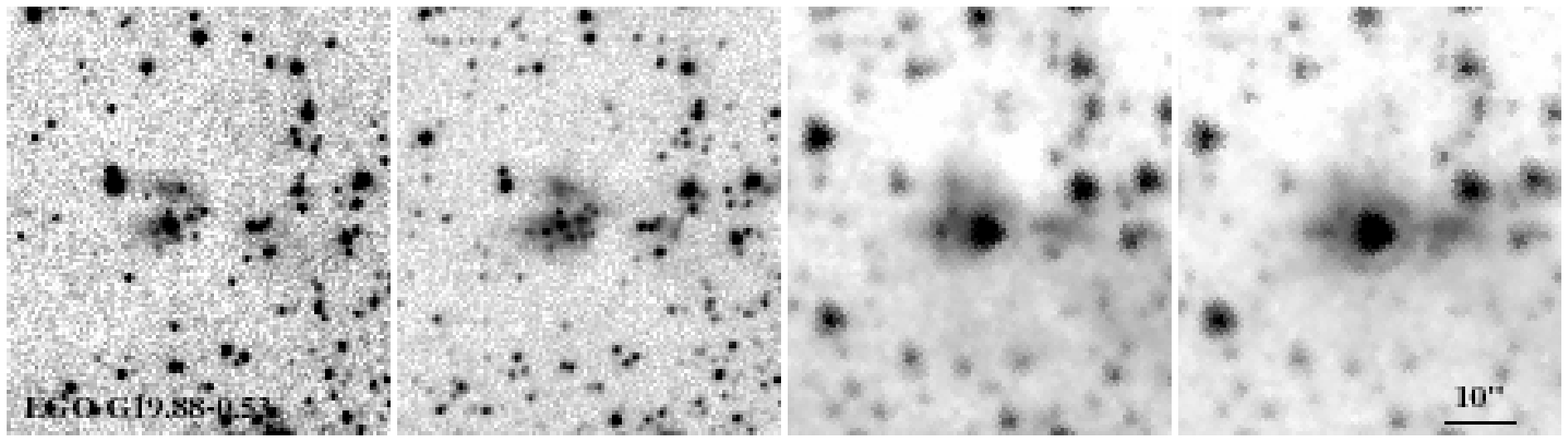}
\includegraphics[angle=0,scale=0.59]{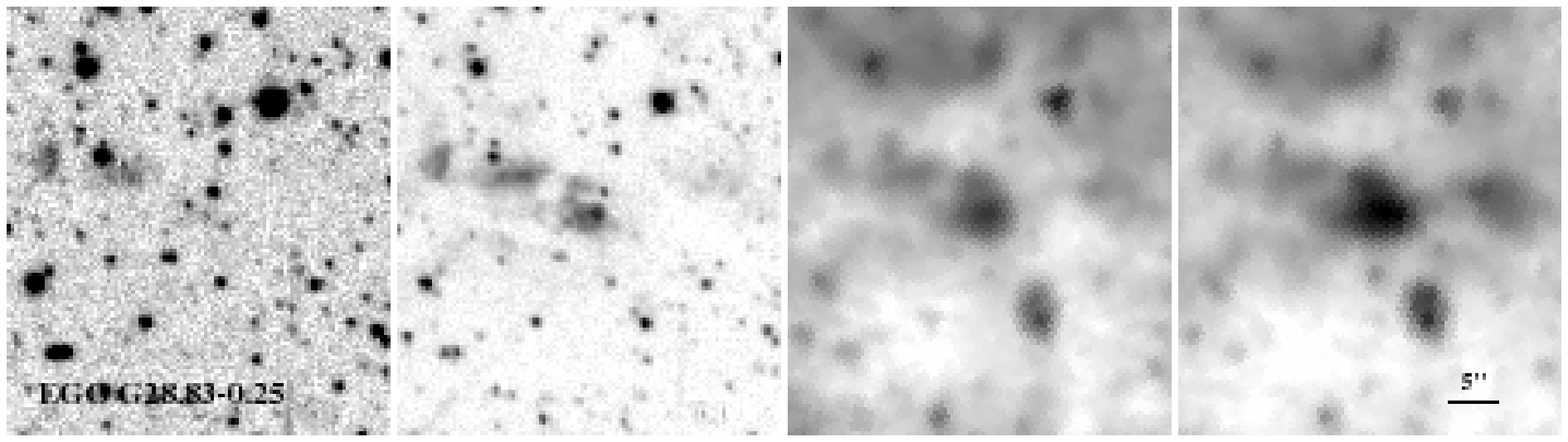}
\includegraphics[angle=0,scale=0.59]{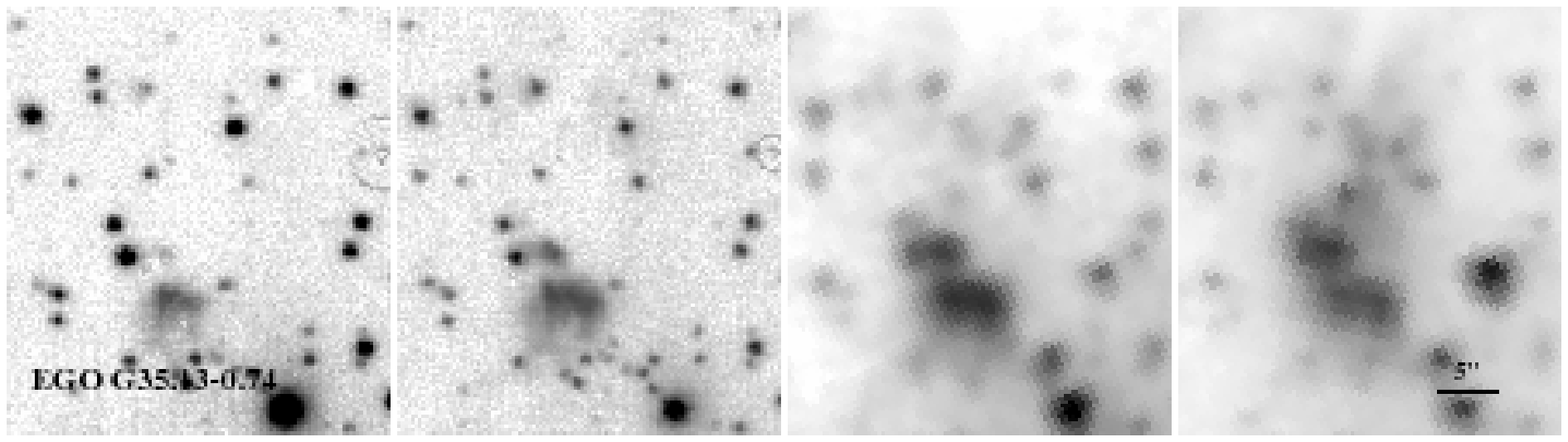}
\includegraphics[angle=0,scale=0.59]{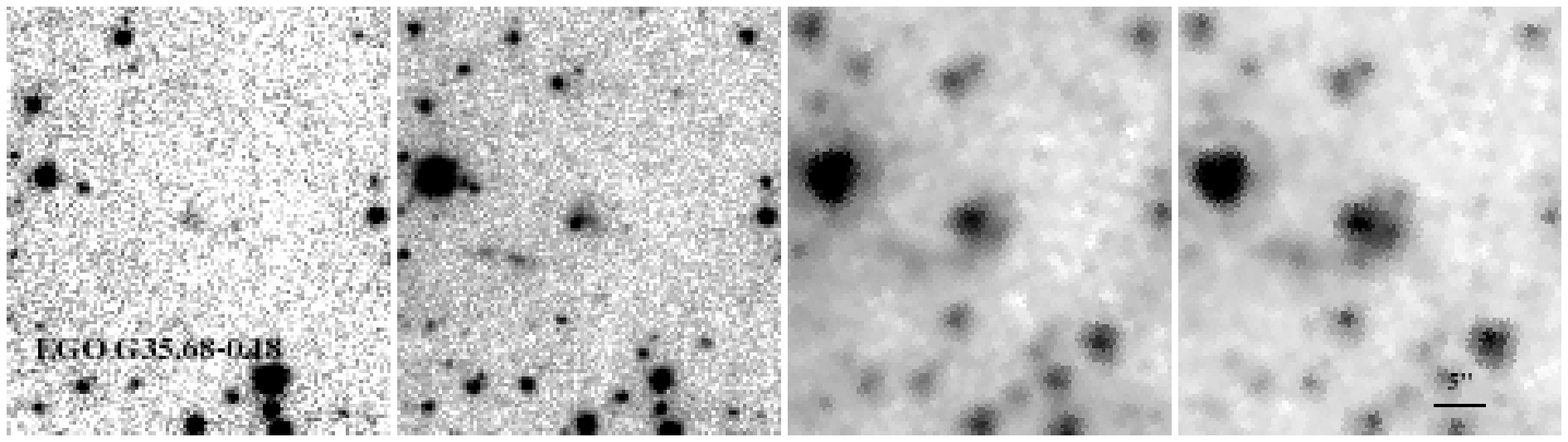}
\includegraphics[angle=0,scale=0.59]{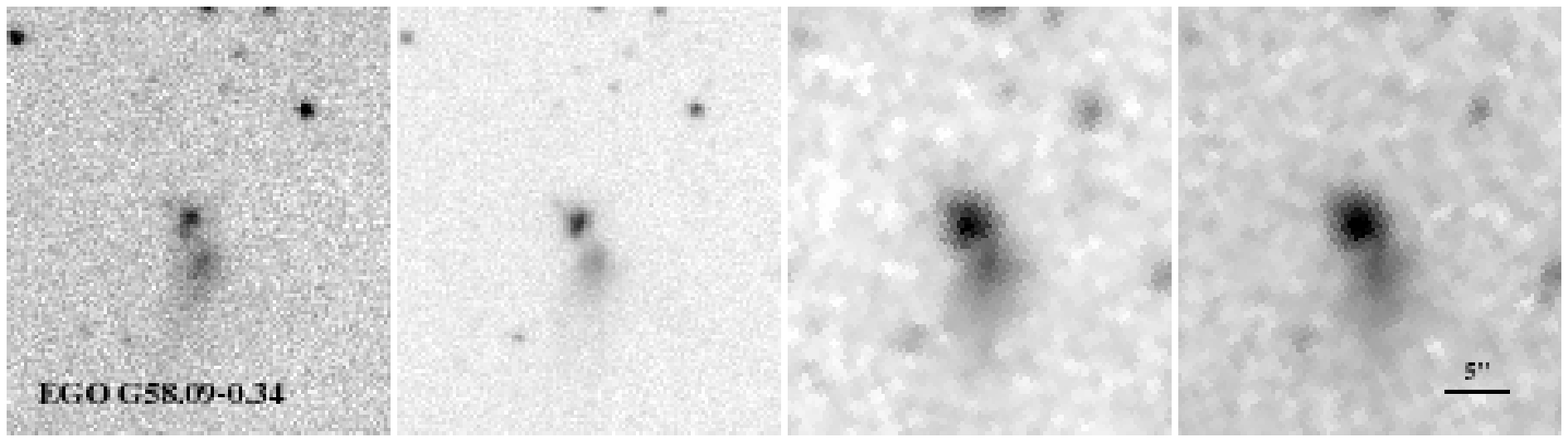}
\caption{EGOs with $H$-band emission; $H$, $K$, 3.6~$\micron$, and 4.5~$\micron$ images in the first, second, third, and fourth columns, respectively.}
\label{fig:hk}
\end{figure}

In the 23 EGO sample, 7 are detected in $H$-band.  Figure~\ref{fig:hk} shows these objects in $H$-, $K$-band, 3.6, and 4.5~$\micron$.  In these figures, the flux distribution gradually changes with wavelength.  In EGO G11.92-0.61, G28.83-0.25 and G35.68-0.18, the extended emission is marginal in the $H$-band, and more apparent at longer wavelengths.  In EGO G12.42+0.50 the $H$-band emission is observed only at the peak of the EGO at 3.6 and 4.5~$\micron$.  The remaining extended components, in particular that to the southwest, are apparent only at 3.6 and 4.5~$\micron$.  In EGO 19.88-0.53, the bright point source is seen at the center only at 3.6 and 4.5~$\micron$.  The simplest explanation of the above characteristics is the effect of extinction at different wavelengths.  %In general, the morphologies of the extended emission gradually changes with wavelength at $H$-band, $K$-band, 3.6, and 4.5~$\micron$.

%For the 12 EGOs with H$_{2}$ outflows, three of them show that the morphology of the H$_{2}$ emission resembles that of the 4.5~$\micron$ emission (EGO G35.68-0.18, G35.83-0.20, and west lobe of G19.88-0.53).  In contrast, the other 9 EGOs with H$_{2}$ outflows show that the morphologies of the $K$ and IRAC-band continuum emission are similar, but they are different from that of H$_{2}$ emission.  Generally, the near-infrared and IRAC-band continuum emission is more compact than the H$_{2}$ emission, and is located around the 24~$\micron$ sources.  However, the H$_{2}$ emission is more extended than the continuum feature, and is seldom associated with the near- and 4.5~$\micron$ emission for the 9 EGOs.  In addition, four EGOs have the orientations of the elongated 4.5~$\micron$ continuum that are different from those of the H$_{2}$ emission (EGO G16.61-0.24, G35.04-0.47, G35.13+0.47, and G35.15+0.80).

\section{DISCUSSION}%4

%In this section, we discuss the results of the 23 EGOs presented in the Appendix.

%Almost all of the EGOs with the $K$-band emission are associated with 24~$\micron$ sources which are likely the origin of the $K$-band scattered emission.  In contrast to the $K$-band, the $H$-band does not contain H$_{2}$ line, so the $H$-band emission is likely to originate from the scattered light from the embedded YSOs.  The morphologies of $H$- and $K$-band are similar, supporting that the near-infrared emission is contributed by the scattered light \citep{hod94,con07,sim09,wal90}.  In addition, the morphology of the near-infrared emission resembles that of the mid-infrared emission as discussed above.  It implies that at least a part of the mid-infrared emission of the EGOs coming from the scatter light from the embedded YSOs.  

\subsection{Morphology of H$_{2}$ Outflow}%4.1

High-mass protostellar outflows show a variety of morphologies, and their driving mechanism is less clear than for low-mass protostellar outflows.  Well known energetic outflows like the Orion BN/KL and DR 21 outflows show a wide opening angle and complicated shock structures \citep[see e.g.,][]{kai00,dav96}. In contrast, some high-mass protostar drive a relatively collimated outflow analogous to that associated with low-mass protostars \citep{var10}.  These tend to be associated with a protostar with spectral types of late O or B, not with the earlier spectral types \citep[see][for review]{beu05,arc07}. %The opening angle of the outflow may represent the evolutionary status of individual high-mass protostars \citep{beu05}.

\citet{beu05} proposed an evolution scenario of outflows from massive stars.  In their scenario, an outflow from a massive star is collimated in their early evolutionary phases.  As it grows by accretion, it reaches the main sequence, associated with an ultra compact \ion{H}{2} region.  The opening angle of the outflow could then be increased by the ionized wind from the massive star.

As mentioned in Section 3.1, the H$_{2}$ emission in most of the EGOs is associated with a collimated outflow.  This suggests that they are relatively young, or their masses are not very high, according to the scenario proposed by \citet{beu05}.  In contrast, EGO G35.20-0.74 exhibits a large opening angle H$_{2}$ outflow.  The hourglass shape outflow is unlikely to be a composite of two collimated outflows at different directions, because the shapes, lengths, and intensities of MHO 2431C and D are similar (Figure~\ref{fig:G35.20-0.74}).  In addition, CO interferometric observations also support that it is a single outflow with large opening angle \citep{bir06}.  The H$_{2}$ outflow is likely to be driven by a YSO which is associated with an ultra compact \ion{H}{2} region \citep{den84} at the outflow base, and it is the only H$_{2}$ outflow with a known \ion{H}{2} region in our sample.  Our results support the evolution scenario of \citet{beu05}.

In our sample, two EGOs are likely to be associated with H$_{2}$ outflows from multiple protostars (EGO G19.88-0.53 and G35.13-0.74).  Three radio sources are located near the outflow base of EGO G19.88-0.53 \citep{zap06}, and EGO G35.13-0.74 is associated with a young star cluster Mercer 14 \citep{fro11b}.

\subsection{Origin of the Extended IRAC Emission Associated with EGOs}%4.2

%As shown in Section 3.1, the morphologies of the extended $K$-band continuum emission and the 4.5~$\micron$ emission are similar.  The $K$-band emission usually peaks around the 24~$\micron$ sources, and becomes fainter farther from the peaks.  The morphologies of the $K$-band emission are similar to those of near-infrared reflection nebulae around YSOs \citep{con07}.

In this section, we discuss the emission mechanisms of those EGOs with near-infrared counterparts, based on the 23 EGOs mentioned in \S3.  Originally, \citet{cyg08} suggested that the extended 4.5~$\micron$ emission of EGOs traces shocked molecular gas in outflows.  However, some studies suggest that the 4.5~$\micron$ emission come from scattered light.  For example, the appearance of NGC~6334~V is similar to that of EGOs in the $Spitzer$ three-color image.  In the 2 $\micron$ polarimetric image, NGC~6334~V shows two highly polarized reflection nebulae (scattered light) coinciding with the 4.5~$\micron$ emission and the bipolar outflow \citep{sim09}.  Another source, EGO G35.20-0.74 known as G35.2N, also shows a highly polarized extended reflection nebulae in $K$-band \citep{wal90}.  These two cases suggest that $K$-band emission represents scattered emission from embedded YSOs, which is consistent with results from \citet{hod94} and \citet{con07}.  IRAS 17527-2439 has an S-shaped H$_{2}$ outflow which may be caused by the precession of a jet \citep{var11}, and its appearance in the $Spitzer$ image is similar to that of EGOs.  Interestingly, the morphologies of the continuum ($K$ and $Spitzer$ bands) and H$_{2}$ emission of IRAS 17527-2439 are different.  The continuum emission in $K$ and $Spitzer$ bands in the direction of the outflow is rotated counterclockwise with respect to the H$_{2}$ outflow, suggesting that the continuum emission emerges from the outflow cavity of the YSO \citep{var11}.  Recently, \citet{sim12} took spectra of EGOs at 5 -- 10~$\micron$ wavelength, and found that the 4.5~$\micron$ emission is not due to H$_{2}$ lines which would be too faint to contribute the IRAC 4.5~$\micron$ emission.

In our sample, only three of the 12 EGOs with H$_{2}$ outflows show a distribution of the H$_{2}$ emission similar to that of the 4.5~$\micron$ emission (EGO G35.68-0.18, G35.83-0.20, and west lobe of EGO G19.88-0.53).  The 4.5~$\micron$ emission of these three EGOs likely comes from H$_{2}$ emission.  The $L$- and $M$-band spectrum of EGO G19.88-0.53 show H$_{2}$ emission \citep{deb10}, supporting this idea.

In contrast, for the remaining 9 EGOs with H$_{2}$ outflows in our sample, we found that the morphologies of the 4.5~$\micron$ and H$_{2}$ emission are different.  The peaks of the 4.5~$\micron$ emission are rarely associated with H$_{2}$ emission, and the 4.5~$\micron$ emission generally appears close to the bases of the outflows (embedded YSOs).  The H$_{2}$ emission is more extended than the 4.5~$\micron$ emission which is usually weak or not detected at the peak of the H$_{2}$ emission.  This suggests that the origins of the H$_{2}$ and 4.5~$\micron$ emission are different.

%{\bf In Figure~\ref{fig:G35.83-0.20}, the H$_{2}$ knots of EGO G35.83-0.20 are associated with the 4.5~$\micron$ emission, so a part of the 4.5~$\micron$ emission may be contributed by the H$_{2}$ emission.  Note that }

For those EGOs with $K$-band emission, the morphologies of the extended $K$-band emission and the 4.5~$\micron$ emission resemble each other, suggesting the same origin.  Furthermore, seven of the sources with extended $K$-band emission also show extended emission in the $H$-band, with flux distributions similar to the $K$-band emission.  This $H$-band emission is not likely to be due to shocked emission, since (1) the flux distribution is significantly different between H$_{2}$ and $K$-band emission; and (2) shocked emission in the $H$-band should be so faint that it is not likely to be observed \citep[e.g.,][]{gre94,eve95}.

%If the 4.5~$\micron$ emission comes from the shocked H$_{2}$ emission, we would expect the morphologies and orientations of the 4.5~$\micron$ and H$_{2}$ emission are similar.  Previous studies showed that the H$_{2}$ outflows are brighter at the shocked regions (especially at the two ends of outflows), but the H$_{2}$ emission is usually weak around driving sources \citep{dav10}.  

%\# of EGOs show that the $K$-band continuum emission is distributed along H$_{2}$ outflows, but is not as extended as the H$_{2}$ emission ().

These observed characteristics can be consistently explained if the extended emission at $H$, $K$, 3.6, and 4.5~$\micron$ is due to scattered continuum \citep{hod94,con07,sim09,wal90}.  Indeed, all of these sources are associated with 24~$\micron$ source(s), i.e., a protostar or multiple protostars, which are the sources of the scattered light.  In Figure~\ref{fig:hk}, the gradual change in flux distribution with wavelength is consistent with extinction at different wavelengths.  In EGO G35.20-0.74, the emission in $K$-band and 4.5~$\micron$ shows a bipolar distribution about the position of the protostar seen at 24~$\micron$. This implies that the scattering occurs in an outflow cavity \citep{wal90}, as in low-mass protostars \citep[e.g.,][]{tob08,sea08}.  %Throughout, we suggest that at least a part of the 4.5~$\micron$ emission from EGOs comes from scattered light from the embedded YSOs.

%If the 4.5~$\micron$ emission comes from shocked gas, then the orientations of the 4.5~$\micron$ and H$_{2}$ emission should be similar.  However, four EGOs show different orientations of the elongated 4.5~$\micron$ continuum and H$_{2}$ emission.  It could be due to (1) precession of H$_{2}$ outflows causing the different orientations between H$_{2}$ and continuum emission \citep[as IRAS 17527-2439,][]{var11}; or (2) scattered light (4.5~$\micron$) from non-uniform ambient gas.  Based on the different orientations between the 4.5~$\micron$ and H$_{2}$ emission and different morphologies discussed above, the implication is that the origins of IRAC-band continuum and H$_{2}$ emission are unlikely to be similar.  

In our 23 EGO sample, only three are consistent with the 4.5~$\micron$ emission coming from the H$_{2}$ emission, and the rest are likely associated with scattered light from the embedded YSOs or mixture of both.  Throughout, we suggest that the 4.5~$\micron$ emission could result from scattered light or/and H$_{2}$ emission.  However, the 4.5~$\micron$ emission itself can not tell which emission mechanism dominates, and we need other wavelength observations (e.g., narrow H$_{2}$- and $K$-band) to distinguish the contributions of these two emission mechanisms.

%Of the 12 EGOs with H$_{2}$ outflows, 11 of them show the $K$-band emission.  Similar to the 4.5~$\micron$ emission, the $K$-band emission also distributes near the bases of the outflows.  However, the H$_{2}$ emission is usually week there, and the distribution H$_{2}$ emission is more extended than that of the $K$-band emission.  %It is probably not due to extinction, as there is $K$-band emission near the bases of these outflows.

%Based on our H$_{2}$ and near-infrared results for those EGOs with near-infrared emission, the mid-infrared emission could be the counterpart of the near-infrared nebulae with the the infrared emission from the embedded YSOs being reflected of the outflow cavity.

\section{CONCLUSIONS}%5

We analyzed the data from H$_{2}$ narrow band, $K$-, and $H$-band observations from UKIRT for 34 EGOs (12 EGOs are preselected and 22 are covered by the UWISH2 survey).  Our narrow band H$_{2}$ observations provide high resolution images ($\sim1\arcsec$) of EGOs, dedicated to shock-excited H$_{2}$ emission, $\upsilon$ =1-0 S(1).  The H$_{2}$ images allow us to determine if these EGOs are associated with H$_{2}$ outflows.  We found that 12 EGOs show H$_{2}$ outflows and four EGOs are H$_{2}$ outflow candidates.  Within this group of 12 EGOs with H$_{2}$ outflows two objects show a chain of H$_{2}$ knotty structures, 5 display extended H$_{2}$ bipolar structures, three have extended H$_{2}$ lobes, and two have a pair of H$_{2}$ knots.  Most of them appear to be associated with a collimated outflow driven by a single protostar.  In contrast, EGO G35.20-0.74 shows an hourglass shape bipolar outflow, and two EGOs (EGO G19.88-0.53 and G35.13-0.74) exhibit multiple outflows.

Of the 12 EGOs with H$_{2}$ outflows, three EGOs show similar distributions between the 4.5~$\micron$ and H$_{2}$ emission, implying similar origins.  For the remaining of the 9 EGOs with H$_{2}$ outflows, however, we found that the peaks of the 4.5~$\micron$ emission are rarely associated with the H$_{2}$ emission.  The H$_{2}$ features are more extended than the continuum features.  %Four EGOs show that the orientations of the elongated 4.5~$\micron$ continuum are different from those of the H$_{2}$ emission.  These results suggest that the origins of the 4.5~$\micron$ and H$_{2}$ emission may be different.

In our sample, we found similar morphologies between the $K$-band and 4.5~$\micron$ emission.  Seven EGOs also exhibit extended emission in $H$-band.  The morphologies of the extended emission gradually changes with wavelength at $H$-band, $K$-band, 3.6, and 4.5~$\micron$.  This can be explained if the emission is associated with scattered continuum, e.g., in the outflow cavity.  The different morphologies at different wavelengths may be attributed to extinction.

\acknowledgments

We thank Drs.~Christopher Davis and Bringfried Stecklum for suggestions and comments.  This research made use of the SIMBAD data base operated at CDS, Strasbourg, France, and NASA's Astrophysics Data System Abstract Service.  MT is supported from National Science Council of Taiwan (Grant No. NSC-100-2112-M-001-007-MY3).

\appendix

\section{Selected Individual Sources}\label{appendix}

\begin{figure}%1
\includegraphics[angle=0,scale=0.8]{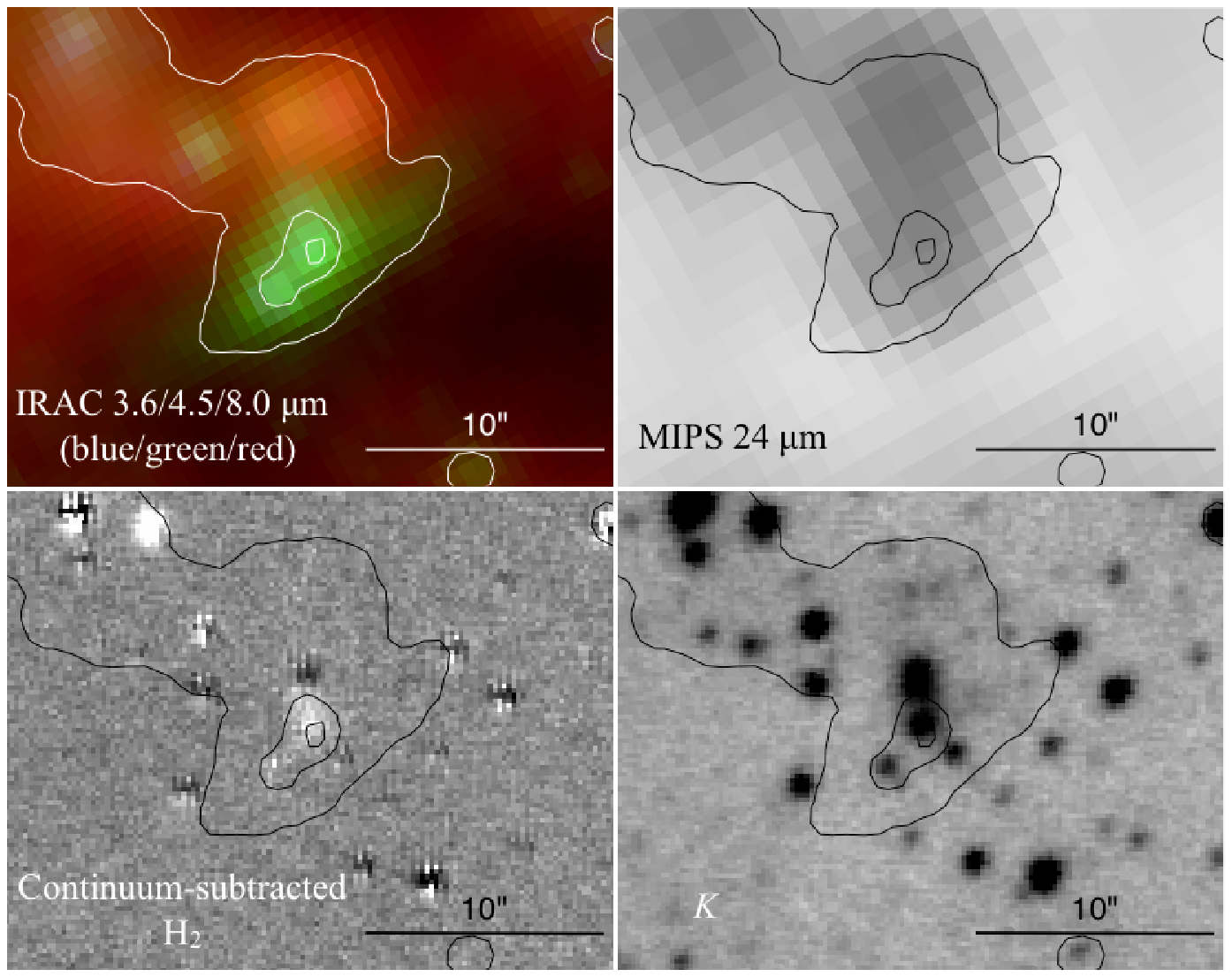}
\caption{Images of EGO G10.34-0.14, upper-left panel: $Spitzer$ IRAC image showing 3.6~$\micron$ (blue), 4.5~$\micron$ (green), and 5.8~$\micron$ (red); upper-right panel: $Spitzer$ MIPS 24~$\micron$ image; lower-left panel: continuum-subtracted H$_{2}$ image; lower-right: GPS $K$-band image.  In addition, 4.5~$\micron$ contours are superposed.  The contour levels are arbitrarily chosen for the best comparisons of the flux distribution between the 4.5~$\micron$ emission and the others.  In the continuum-subtracted H$_{2}$ image, compact features with a combination of positive and negative valued features are the residual of continuum subtraction of point sources.  In addition to those features, values lower than the background are observed at the peak of the EGO and $\sim10\arcsec$ northeast.  This is due to continuum whose flux ratio between the H$_{2}$ narrow band and $K$-band filters are different from that of foreground stars.  The resolutions of IRAC, MIPS, and continuum-subtracted H$_{2}$ images are $\sim$2$\arcsec$, 6$\arcsec$, and less than 1$\arcsec$, respectively.}
% contour levels: 20, 100, 180
\label{fig:G10.34-0.14}
\end{figure}

%2-2, f1
{\bf EGO G10.34$-$0.14} (Figure~\ref{fig:G10.34-0.14}).  The 24~$\micron$ image is marginally extended in the EGO region, presumably due to the presence of multiple sources.  No H$_{2}$ feature is detected in the continuum-subtracted H$_{2}$ image.  Two negative valued features are seen in the continuum-subtracted H$_{2}$ image.  One is located at the 4.5~$\micron$ peak flux position of the EGO, the other $\sim10\arcsec$ northeast of the EGO.  In the $K$-band image, both extended and point sources are seen at the position of the EGO, and coincide with the 24~$\micron$ source.
%\citet{cyg09} found that the spatial relationship and velocity distribution of the Class I methanol masers are complex around the EGO.

\begin{figure}%2
\includegraphics[angle=0,scale=0.55]{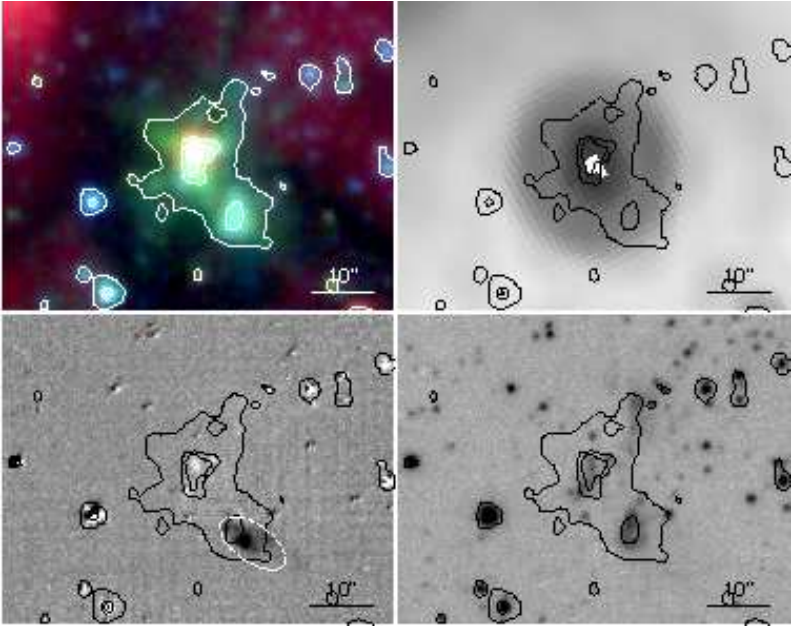}
\caption{Same as Figure~\ref{fig:G10.34-0.14}, but for EGO G11.92-0.61.  The dashed ellipse illustrates a lobe of H$_{2}$ outflow (MHO 2303).  The white part in the 24~$\micron$ source is saturated.}
% contour levels: 10, 80, 150
\label{fig:G11.92-0.61}
\end{figure}

%1-1, f2
{\bf EGO G11.92$-$0.61} (Figure~\ref{fig:G11.92-0.61}).  A molecular outflow along the NE-SW direction was discovered by \citet{cyg11}.  The EGO exhibits two peaks in the NE-SW direction with a separation of $\sim12\arcsec$ in the 4.5~$\micron$ contour map.  A bright 24~$\micron$ source is associated with the NE 4.5~$\micron$ peak of the EGO.  H$_{2}$ emission is identified near the SW peak of the EGO, and extends to the SW direction.  This flow component is presumably associated with the blue lobe of the molecular outflow discovered by \citet{cyg11}.  The distribution of H$_{2}$ emission differs from that of the $K$-band and 4.5~$\micron$ continuum, while the distributions of the $K$-band and 4.5~$\micron$ continuum match more closely.  In the $K$-band image, there are two extended sources at the two peaks of the 4.5~$\micron$ emission.

\begin{figure}%3
\includegraphics[angle=0,scale=0.55]{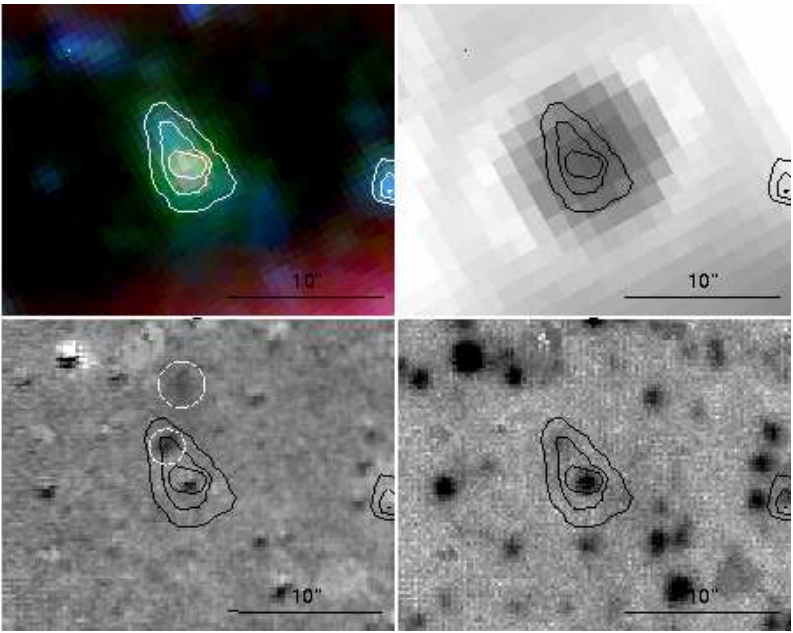}
\caption{Same as Figure~\ref{fig:G10.34-0.14}, but for EGO G12.02-0.21.  One H$_{2}$ knot (dashed circle near the 4.5~$\micron$ peak) and one possible H$_{2}$ knot (northern dashed circle) are detected.}
% contour levels: 10, 20, 30
\label{fig:G12.02-0.21}
\end{figure}

%1-2, f3
{\bf EGO G12.02$-$0.21} (Figure~\ref{fig:G12.02-0.21}).  EGO G12.02-0.21 is associated with a 24~$\micron$ source located at the 4.5~$\micron$ peak.  There is a faint H$_{2}$ knot and another possible H$_{2}$ knot located 3$\arcsec$ and 6$\arcsec$ north of the 24~$\micron$ source, respectively.  The $K$-band image shows a continuum source at the 4.5~$\micron$ peak of the EGO.

\begin{figure}%4
\includegraphics[angle=0,scale=0.55]{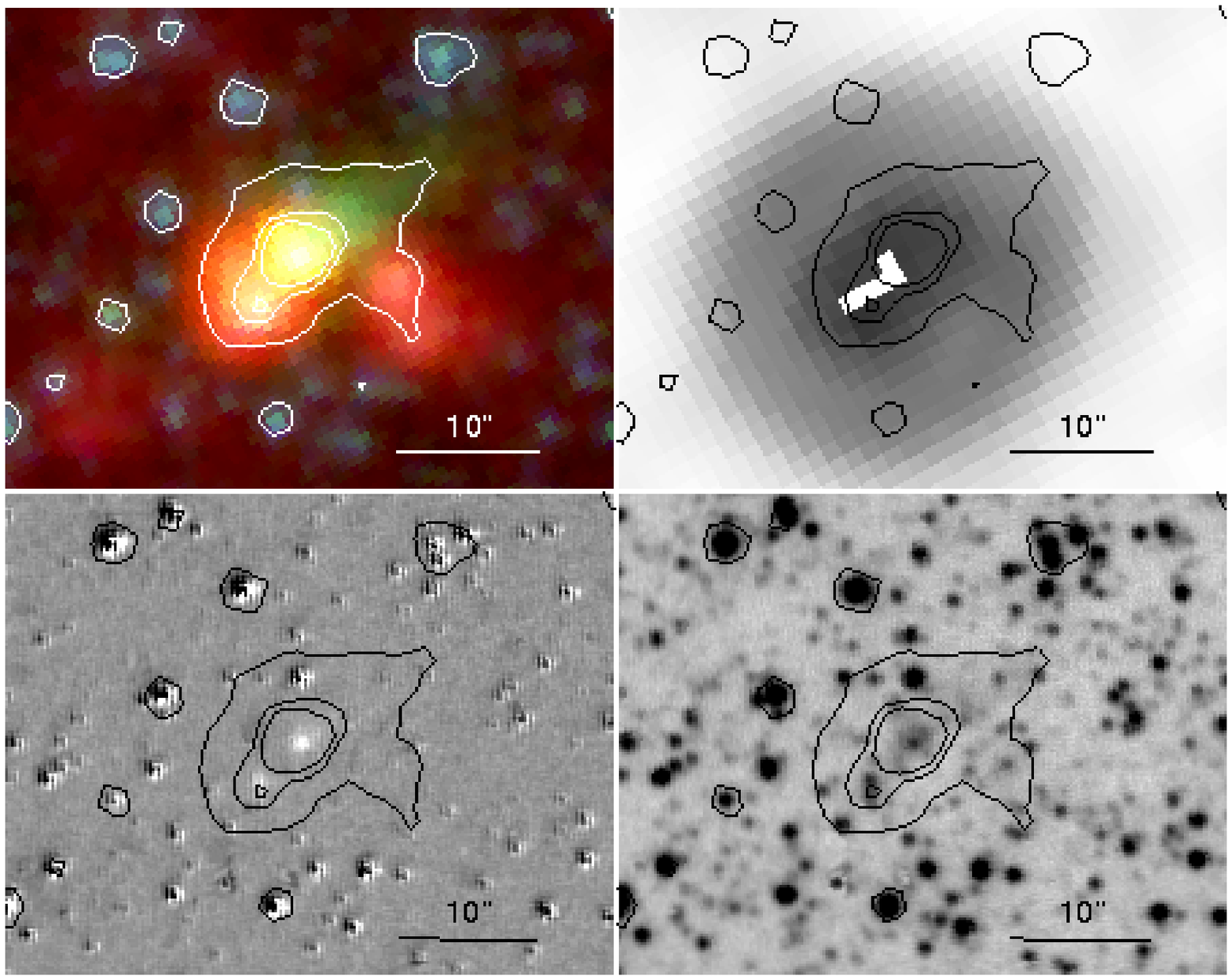}
\caption{Same as Figure~\ref{fig:G10.34-0.14}, but for EGO G12.20-0.03.  The white part in the 24~$\micron$ source is saturated.}
\label{fig:G12.20-0.03}
%contour levels: 12, 62, 112
\end{figure}

%4-1, f4
{\bf EGO G12.20$-$0.03} (Figure~\ref{fig:G12.20-0.03}).  There are two barely resolved 24~$\micron$ sources with a 5$\arcsec$ separation associated with 4.5~$\micron$ peaks in the NW-SE direction.  The SE 4.5~$\micron$ peak coincides with an ultra-compact \ion{H}{2} region, [WBH2005] G012.199-0.034 \citep{urq09}.  In the continuum-subtracted H$_{2}$ image, no H$_{2}$ feature is detected.  We identified two continuum sources in $K$-band image at the two peaks of the 4.5~$\micron$ emission, seen as negative valued features in the continuum-subtracted H$_{2}$ image.

\begin{figure}%5
\includegraphics[angle=0,scale=0.55]{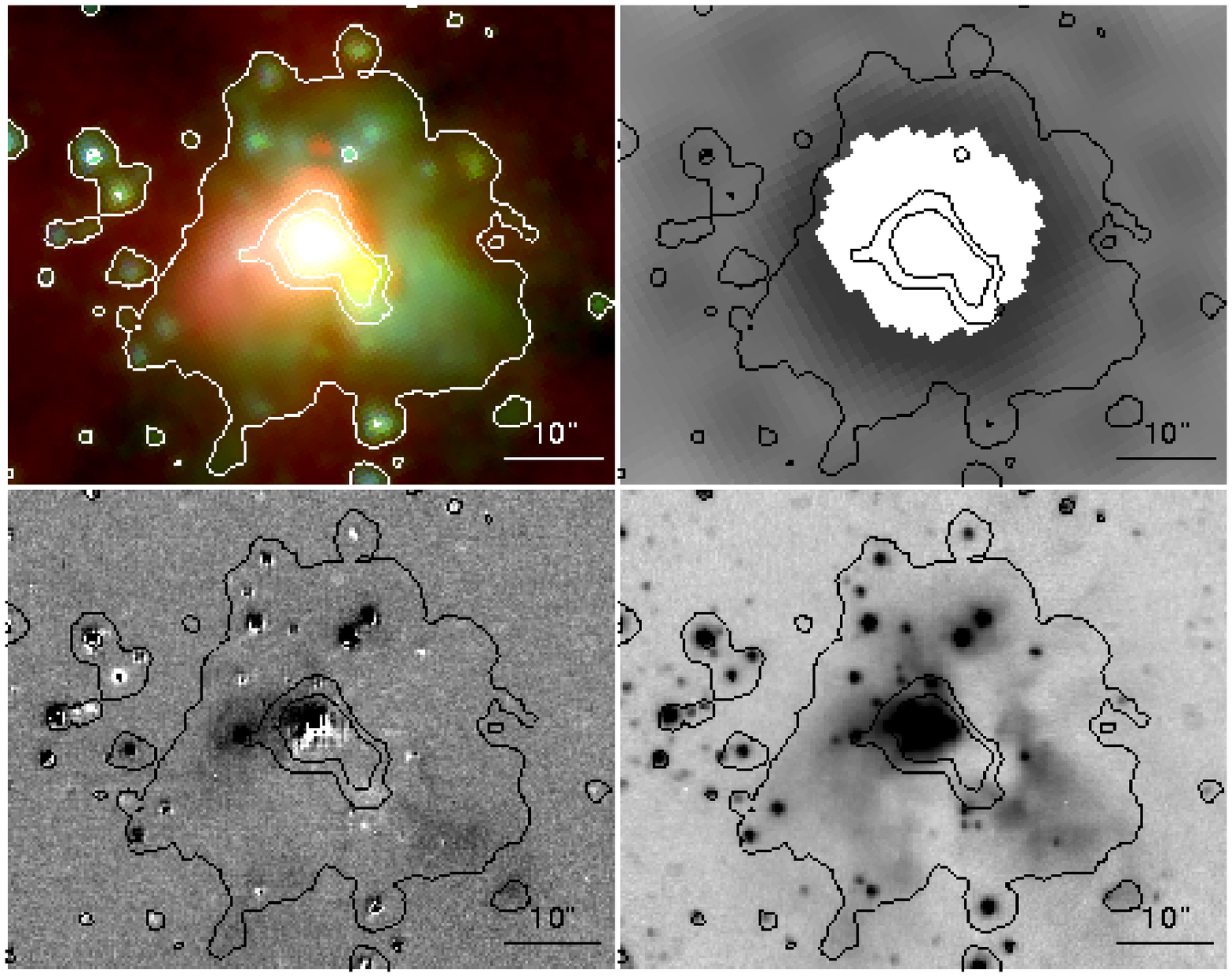}
\caption{Same as Figure~\ref{fig:G10.34-0.14}, but for EGO G12.42+0.50.  The white part in the 24~$\micron$ source is saturated.}
% contour levels: 10, 100, 190
\label{fig:G12.42+0.50}
\end{figure}

%4-2, f5
{\bf EGO G12.42$+$0.50} (Figure~\ref{fig:G12.42+0.50}).  EGO G12.42+0.50 is associated with a very bright 24~$\micron$ source which saturates the detector.  The 24~$\micron$ source coincides with an \ion{H}{2} region, [UHP2009] VLA G012.4180+00.5038 \citep{urq09}, and is located at the peak of 4.5~$\micron$ emission.  There is extended emission in the continuum-subtracted H$_{2}$ image.  This may be due to residuals from the 
continuum subtraction, rather than true H$_{2}$ emission.  In the $K$-band image, there is an extended $K$-band source at the position of the 24~$\micron$ source.  The location of the brightest $K$-band flux coincide in the 4.5~$\micron$ and $K$-band images.

\begin{figure}%6
\includegraphics[angle=0,scale=0.55]{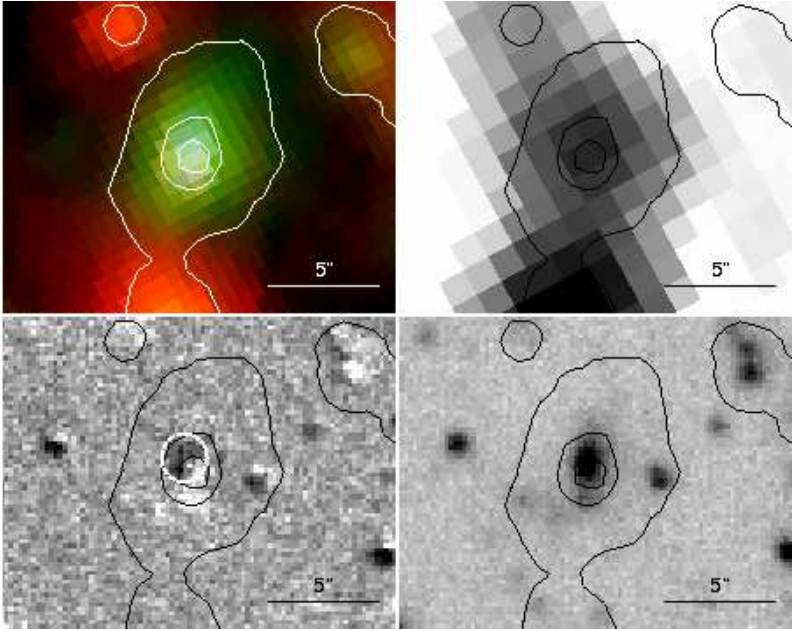}
\caption{Same as Figure~\ref{fig:G10.34-0.14}, but for EGO G12.91-0.03.  The dashed circle marks the position of a possible H$_{2}$ knot.}
%contour levels: 10, 80, 150
\label{fig:G12.91-0.03}
\end{figure}

%1-3, f6
{\bf EGO G12.91$-$0.03} (Figure~\ref{fig:G12.91-0.03}).  EGO G12.91-0.03 is associated with a 24~$\micron$ source at the peak of the 4.5~$\micron$ emission.  There is a possible knot in the H$_{2}$ emission 1$\arcsec$ to the north of the peak in the 4.5~$\micron$ emission.  The $K$-band image shows an extended continuum source very close to the 4.5~$\micron$ peak and the 24~$\micron$ source.

\begin{figure}%7
\includegraphics[angle=0,scale=0.55]{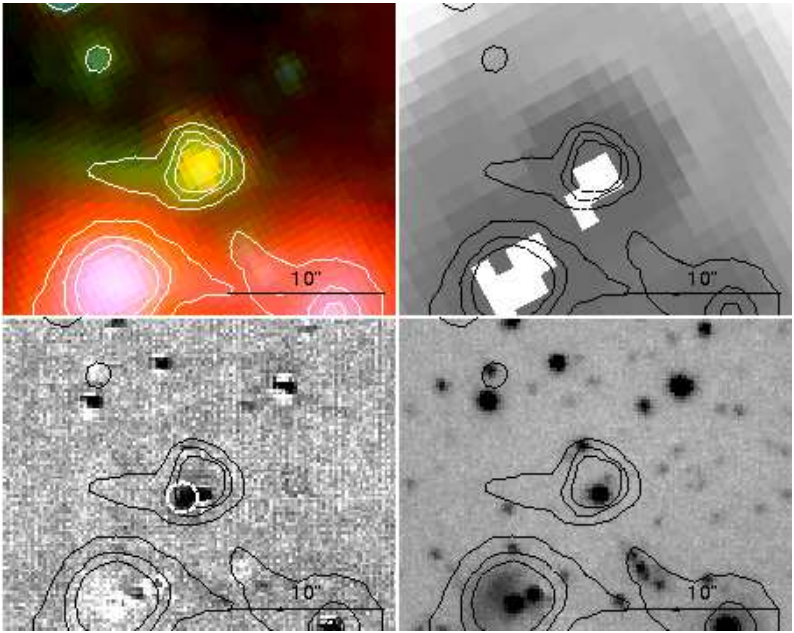}
\caption{Same as Figure~\ref{fig:G10.34-0.14}, but for EGO G16.59-0.05.  The dashed circle marks the position of a possible H$_{2}$ knot.  The white part in the 24~$\micron$ source is saturated.}
%contour levels: 20, 40, 60
\label{fig:G16.59-0.05}
\end{figure}

%2-3, f7
{\bf EGO G16.59$-$0.05} (Figure~\ref{fig:G16.59-0.05}).  EGO G16.59-0.05 is associated with IRAS 18182-1433, a star-forming region with radio free-free emission \citep{zap06}.  \citet{zap06} suggested that this emission is either due to an optically thin \ion{H}{2} region or a thermal jet (or stellar wind).  EGO G16.59-0.05 is associated with a 24~$\micron$ source coinciding with the peak of the 4.5~$\micron$ emission.  The H$_{2}$ knot in the continuum-subtracted H$_{2}$ image, the $K$-band source, and the 4.5~$\micron$ source are close to each other, but not exactly coincident.  

\begin{figure}%8
\includegraphics[angle=0,scale=0.55]{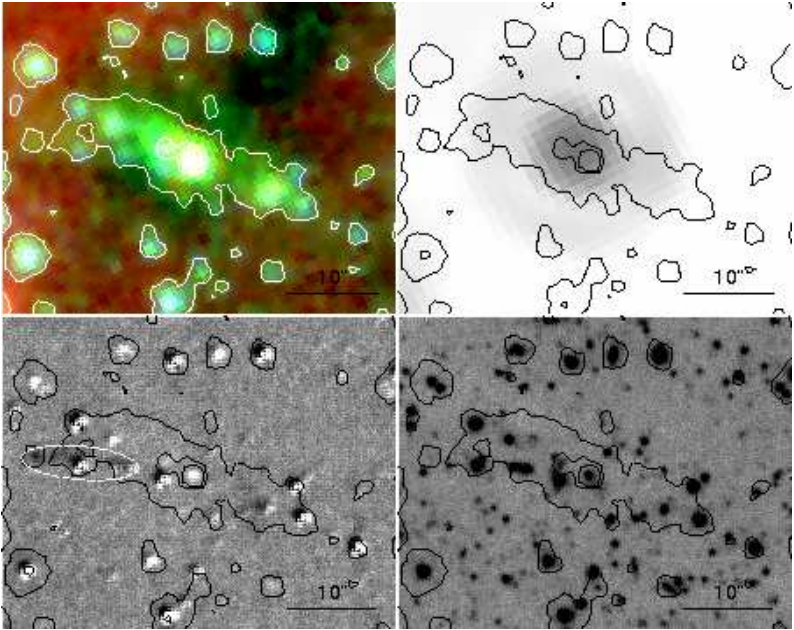}
\caption{Same as Figure~\ref{fig:G10.34-0.14}, but for EGO G16.61-0.24.  The dashed ellipse illustrates a lobe of an H$_{2}$ outflow (MHO 2243).}
%contour levels: 5, 35, 65
\label{fig:G16.61-0.24}
\end{figure}

%1-6, f8
{\bf EGO G16.61$-$0.24} (Figure~\ref{fig:G16.61-0.24}).  EGO G16.61-0.24 is associated with a 24~$\micron$ source located near the peak of the 4.5~$\micron$ emission.  In the continuum-subtracted H$_{2}$ image, there is a faint H$_{2}$ lobe to the east; the tip of the H$_{2}$ outflow may be associated with faint 4.5~$\micron$ emission but the latter detection is not convincing.  The spatial distributions of the H$_{2}$ outflow and bright 4.5~$\micron$ emission in the east side are different, with position angles of 80$\arcdeg$ and 70$\arcdeg$, respectively.  The continuum-subtracted H$_{2}$ image shows negative valued features at the peak of the 4.5~$\micron$ emission.  There is a $K$-band extended source at the position of the 24~$\micron$ source.

%1-8
%{\bf EGO G18.89$-$0.47}. EGO G18.89-0.47.  scatter, methanol masers \citep{cyg09} no detection

\begin{figure}%9
\includegraphics[angle=0,scale=0.55]{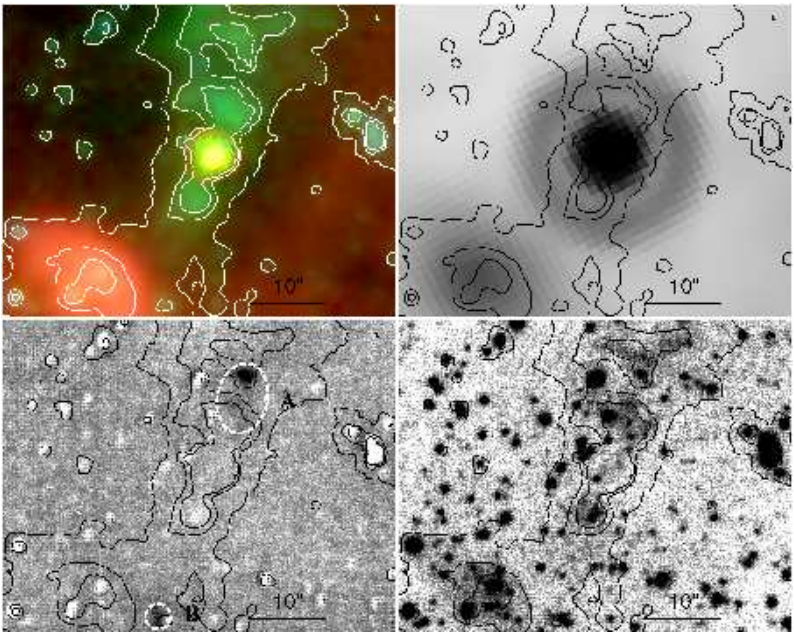}
\caption{Same as Figure~\ref{fig:G10.34-0.14}, but for EGO G19.01-0.03.  The dashed circle and ellipse mark a pair of H$_{2}$ knots (MHO 2244A--B).}
%contour levels: 10, 20, 30
\label{fig:G19.01-0.03}
\end{figure}

%1-9, f9
{\bf EGO G19.01$-$0.03} (Figure~\ref{fig:G19.01-0.03}).  EGO G19.01-0.03 is associated with a 24~$\micron$ source at the 4.5~$\micron$ peak.  The EGO is elongated in the 4.5~$\micron$, and there is a group of 44~GHz methanol masers distributed along the EGO \citep{cyg09}.  A molecular outflow discovered by \citet{cyg11} has a morphology similar to that of the 4.5~$\micron$ emission.  We detected two H$_{2}$ knots near the two lobes of the molecular outflow, and the northern H$_{2}$ knot coincides with the position of the 44~GHz methanol masers.  The symmetric distribution of the two H$_{2}$ knots with respect to the 24~$\micron$ source suggests a bipolar outflow.  The $K$-band extended emission is distributed to the both sides of the outflow.  The morphology of the 4.5~$\micron$ emission is similar to that of the $K$-band continuum, but is different from that of the H$_{2}$ emission.  

\begin{figure}%10
\includegraphics[angle=0,scale=0.55]{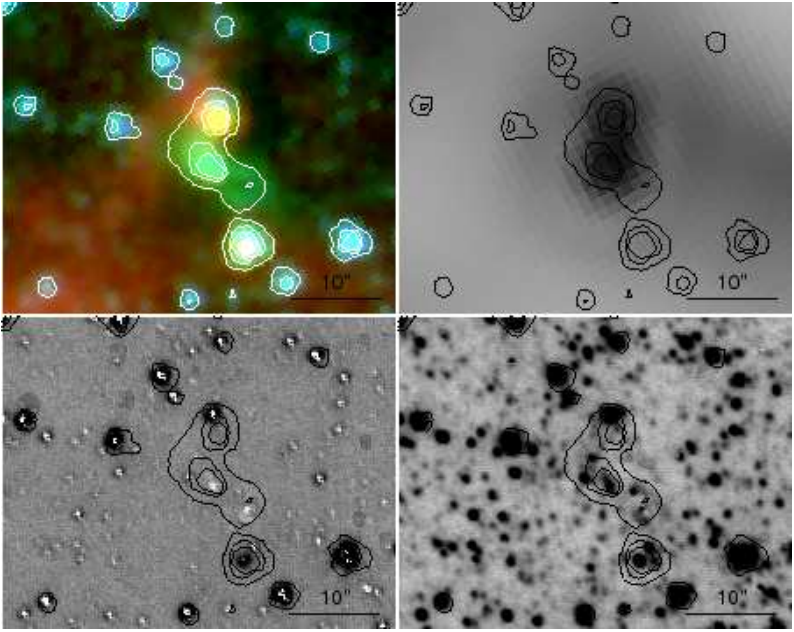}
\caption{Same as Figure~\ref{fig:G10.34-0.14}, but for EGO G19.61-0.12.}
%contour levels: 
\label{fig:G19.61-0.12}
\end{figure}

%2-5, f10
{\bf EGO G19.61$-$0.12} (Figure~\ref{fig:G19.61-0.12}).  In the 4.5~$\micron$ contour map, there are two peaks, each associated with a 24~$\micron$ source.  The southern peak is associated with the EGO identified by \citet{cyg08}.  In the continuum-subtracted H$_{2}$ image, EGO G19.61-0.12 is not associated with any H$_{2}$ feature.  There are two $K$-band sources exhibiting negative valued features in the continuum-subtracted H$_{2}$ image.  One is associated with the EGO (southern 4.5~$\micron$ peak), and the other is located to the S-E direction with a $\sim5\arcsec$ separation.  %The $K$-band source associated with the EGO exhibits extended emission, and coincides with the 24~$\micron$ source.  The extended $K$-band emission is observed at the brightest peak of the 4.5~$\micron$ emission and is elongated in the NE-SW direction.

%A $K$-band source exhibits negative value (continuum emission) in the continuum-subtracted H$_{2}$ image is located at the southern 4.5~$\micron$ peak.

\begin{figure}%11
\includegraphics[angle=0,scale=0.75]{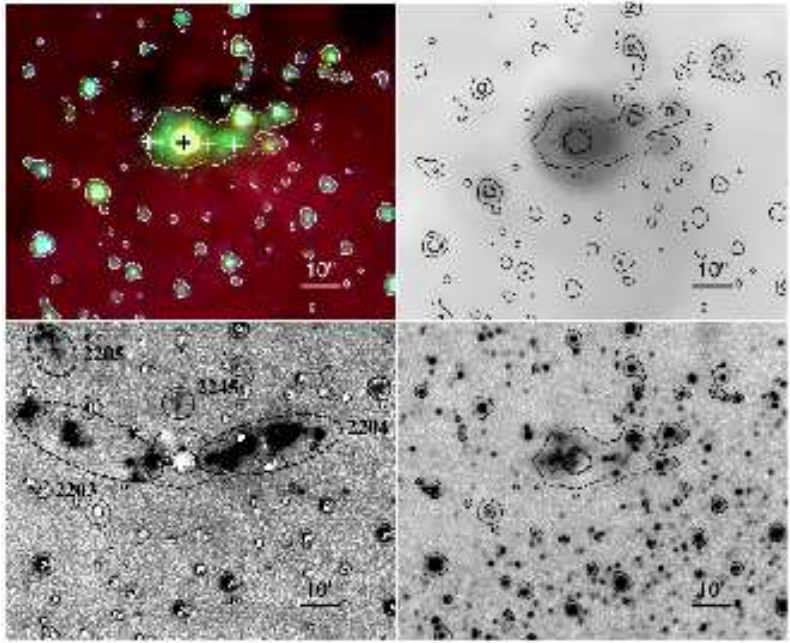}
\caption{Same as Figure~\ref{fig:G10.34-0.14}, but for EGO G19.88-0.53.  The dashed circle and ellipse mark the MHO numbers.}
%contour levels: 10, 60, 110
\label{fig:G19.88-0.53}
\end{figure}

%1-10, f11
{\bf EGO G19.88$-$0.53} (Figure~\ref{fig:G19.88-0.53}).  EGO G19.88-0.53 is related to the IRAS 18264-1152 star-forming region.  The crosses in the GLIMPSE image show positions where \citet{deb10} took spectra in the 3-6~$\micron$ range.  These authors show that the 4.5~$\micron$ emission at the black cross (i.e., the position of a massive YSO) is due to continuum, while it is due to H$_{2}$ emission at the white crosses.  A 24~$\micron$ source is at the peak of the 4.5~$\micron$ emission.  In our image, EGO G19.88-0.53 shows a clear bipolar H$_{2}$ outflow with an angular scale of $\sim1.6\arcmin$ in the E-W direction.  H$_{2}$ knots are also located further out to the northeast and north of the EGO, probably due to multiple outflow sources.  These are marked as MHO 2203, 2204, 2205, and 2245 in the figure.  All except MHO 2245 have been identified in previous narrow-band H$_{2}$ observations by \citet{var10}.  The position of the H$_{2}$ emission close to the driving source matches the diffuse components at 4.5~$\micron$.  The 4.5~$\micron$ and $K$-band emission are not as extended as the H$_{2}$ emission.  The $K$-band image shows a relatively bright extended component at the base of the eastern lobe.  There are also some faint extended component in the $K$-band image in the west side of the protostar(s).

%MHO~2203, 2204, and 2205.

\begin{figure}%12
\includegraphics[angle=0,scale=0.55]{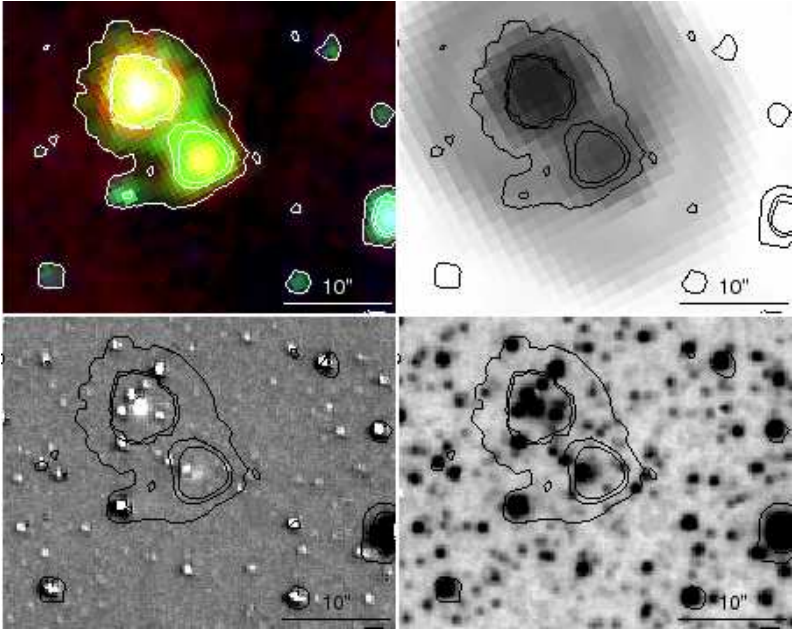}
\caption{Same as Figure~\ref{fig:G10.34-0.14}, but for EGO G20.24+0.07.}
%contour levels: 10, 30, 50
\label{fig:G20.24+0.07}
\end{figure}

%4-6, f12
{\bf EGO G20.24$+$0.07} (Figure~\ref{fig:G20.24+0.07}).  There are two sources seen in the GLIMPSE image; the central source is associated with the EGO \citep{cyg08}, while the other source is located at the north east.  Both sources show strong 24~$\micron$ emission.  No H$_{2}$ emission is seen in our continuum-subtracted H$_{2}$ image, however negative valued features due to continuum emission are associated with both 4.5~$\micron$ sources.  In the $K$-band image, there is an extended source around the 24~$\micron$ source associated with the EGO.

\begin{figure}%13
\includegraphics[angle=0,scale=0.55]{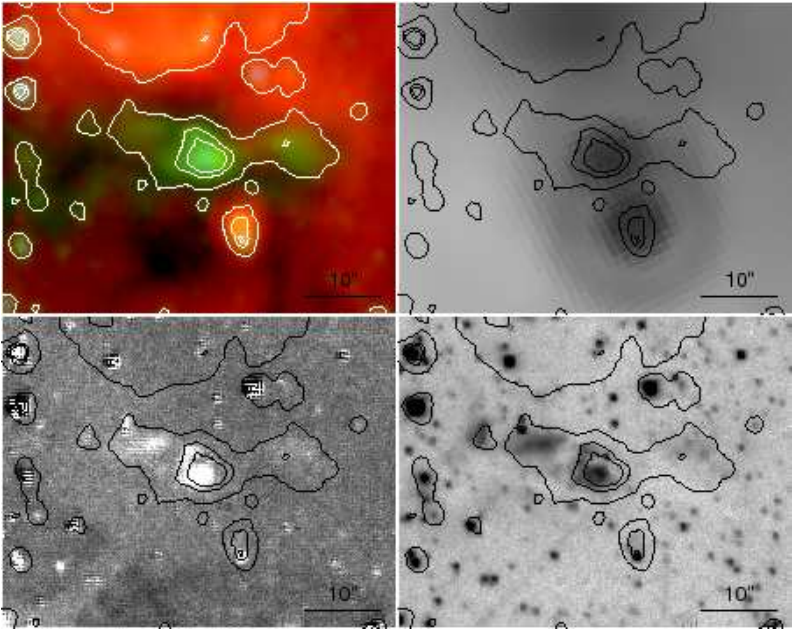}
\caption{Same as Figure~\ref{fig:G10.34-0.14}, but for EGO G28.83-0.25.}
%contour levels: 14, 44, 74
\label{fig:G28.83-0.25}
\end{figure}

%1-19, f13
{\bf EGO G28.83$-$0.25} (Figure~\ref{fig:G28.83-0.25}).  EGO G28.83-0.25 is associated with a 24~$\micron$ source at the peak of the 4.5~$\micron$ emission.  No H$_{2}$ emission is seen in the continuum-subtracted H$_{2}$ image, however negative valued features due to continuum emission are seen, with a morphology similar to the 4.5~$\micron$ emission.  Diffuse $K$-band emission is observed at the peak of the EGO and a faint feature $\sim10\arcsec$ east.

\begin{figure}%14
\includegraphics[angle=0,scale=0.55]{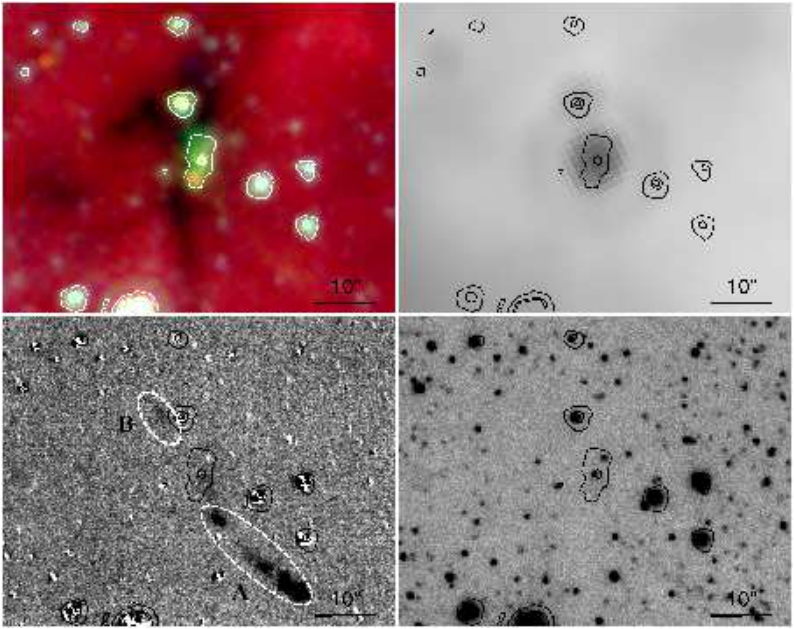}
\caption{Same as Figure~\ref{fig:G10.34-0.14}, but for EGO G35.04-0.47.  The dashed ellipses identify the H$_{2}$ outflow (MHO 2429A--B).}
%contour levels: 10, 40, 70
\label{fig:G35.04-0.47}
\end{figure}

%1-22, f14
{\bf EGO G35.04$-$0.47} (Figure~\ref{fig:G35.04-0.47}).  EGO G35.04-0.47 has a 24~$\micron$ source at the peak of the 4.5~$\micron$ emission.  Its GLIMPSE image shows that it is extended along an N-S axis.  However, our continuum-subtracted H$_{2}$ image shows an H$_{2}$ outflow along NE-SW axis which is not detected in the 4.5~$\micron$ image.  The SW lobe of the H$_{2}$ outflow is much brighter than the NE one which can be explain if the NE lobe is obscured by the IRDC.  The morphology of the H$_{2}$ emission is more extended than that of the 4.5~$\micron$ emission.  A faint $K$-band source is observed at the 4.5~$\micron$ peak (and the 24~$\micron$ source).

\begin{figure}%15
\includegraphics[angle=0,scale=0.55]{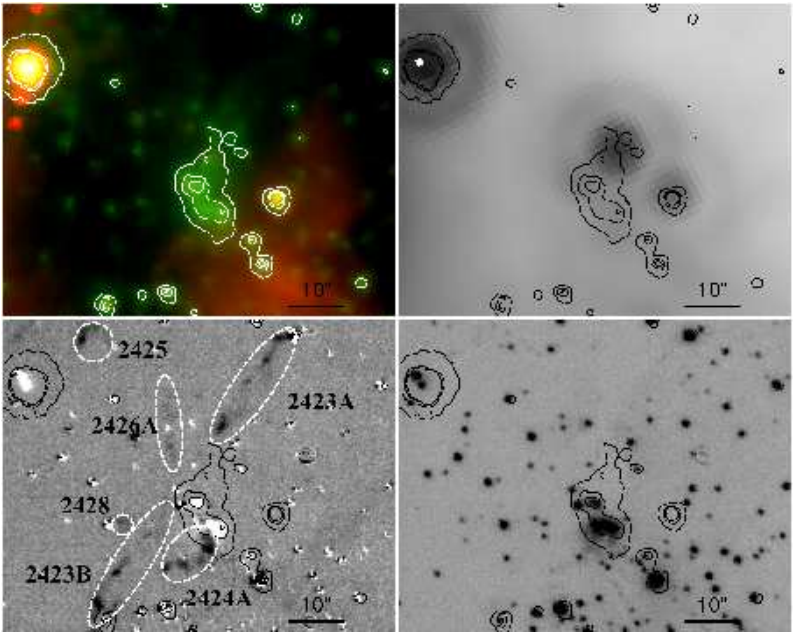}
\caption{Same as Figure~\ref{fig:G10.34-0.14}, but for EGO G35.13-0.74.  The dashed ellipses mark the position of H$_{2}$ features, and their MHO numbers are labeled.}
%contour levels: 30, 80, 130
\label{fig:G35.13-0.74}
\end{figure}

%1-23, f15
{\bf EGO G35.13$-$0.74} (Figure~\ref{fig:G35.13-0.74}).  EGO G35.13-0.74 is likely associated with the embedded cluster [MCM2005b] 14 \citep[Mercer 14,][]{mer05}.  There is a 24~$\micron$ source close to the EGO, but it is not located at the peak of the 4.5~$\micron$ emission.  The continuum-subtracted H$_{2}$ image shows a well collimated H$_{2}$ outflow around the EGO in the NW-SE direction, and also a number of H$_{2}$ knots surrounding it.  The 4.5~$\micron$ image shows a few possible faint counterparts for these H$_{2}$ knots.  In contrast, the 4.5~$\micron$ and $K$-band images only show extended emission, with similar morphologies.  For further information regarding H$_{2}$ knots (MHO 2423-2428) and their possible driving sources, see \citet{fro11b} in detail.

\begin{figure}%16
\includegraphics[angle=0,scale=0.55]{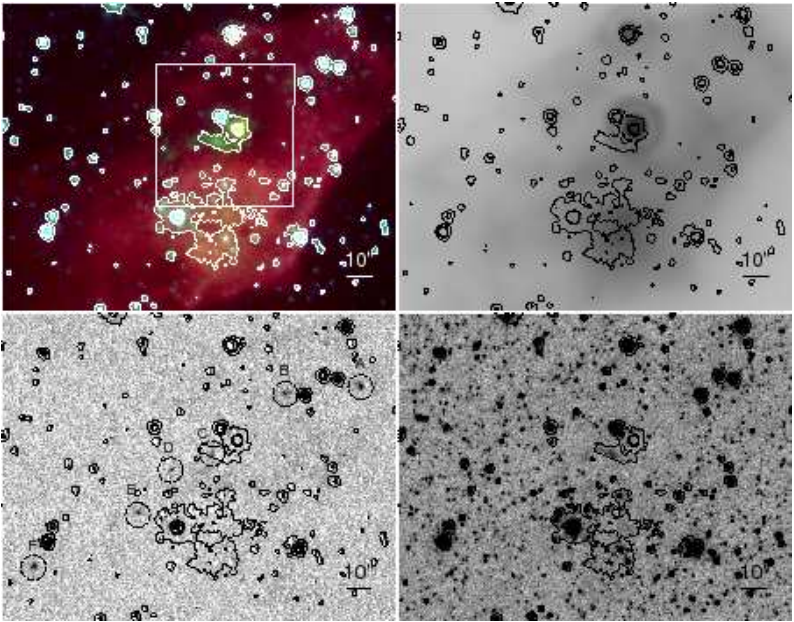}
\caption{Same as Figure~\ref{fig:G10.34-0.14}, but for EGO G35.15+0.80.  The lowest contour around the EGO roughly represents its shape.  The dashed circles illustrate the H$_{2}$ knots of the H$_{2}$ outflow (NHO 2430A--F).  The white box is the region which will be shown in Figure~\ref{fig:G35.15+0.80b}.}
%contour levels: 7, 37, 67
\label{fig:G35.15+0.80}
\end{figure}

\begin{figure}%16b
\includegraphics[angle=0,scale=0.55]{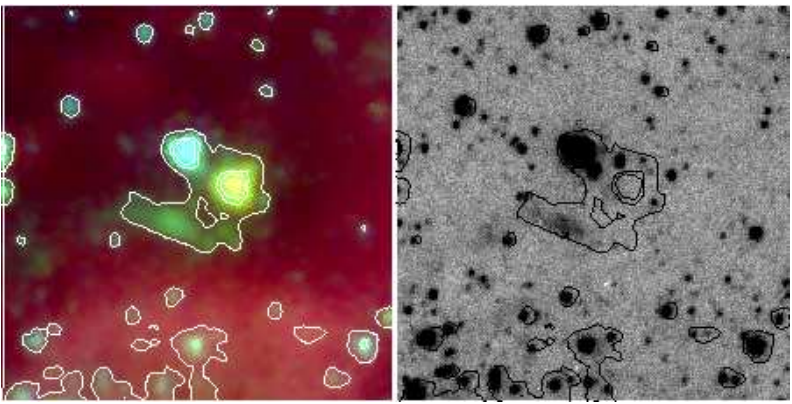}
\caption{$Spitzer$ and $K$-band images of EGO G35.15+0.80.  The lowest contour around the EGO represents its approximate shape.}
%contour levels: 7, 37, 67
\label{fig:G35.15+0.80b}
\end{figure}

%1-24, f16
{\bf EGO G35.15$+$0.80} (Figure~\ref{fig:G35.15+0.80}).  EGO G35.15+0.80 is associated with a 24~$\micron$ source at the 4.5~$\micron$ peak.  The EGO is at the center of the image and is extended in the NE-SW direction.  In the H$_{2}$ image, EGO G35.15+0.80 shows a set of aligned H$_{2}$ knots in the NW-SE direction, different from the orientation of the 4.5~$\micron$ emission.  The 4.5~$\micron$ image shows a few possible faint counterparts to these H$_{2}$ knots.  The $K$-band extended emission is located $\sim$10$\arcsec$ SE of the 24~$\micron$ source.  The morphology of 4.5~$\micron$ emission resembles that of $K$-band continuum emission, except there is no $K$-band point source at the position of the 24~$\micron$ source (Figure~\ref{fig:G35.15+0.80b}).

\begin{figure}%17+1
\includegraphics[angle=0,scale=0.42]{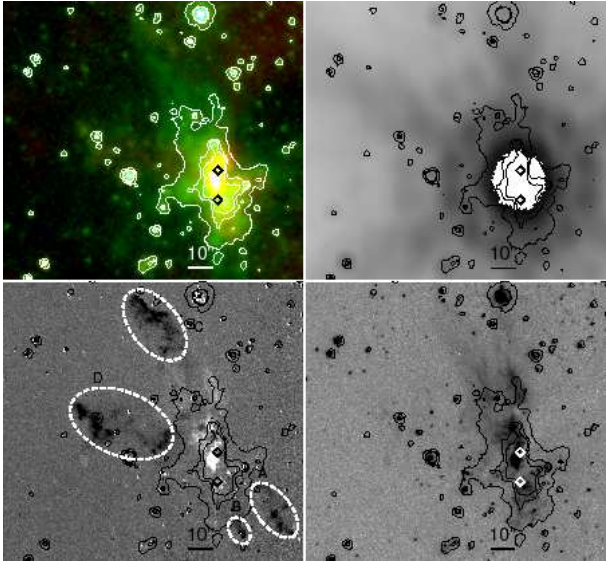}
\caption{Same as Figure~\ref{fig:G10.34-0.14}, but for EGO G35.20-0.74.  Two diamonds label the positions of the ultra compact \ion{H}{2} regions \citep{den84}.  The dashed ellipse marks the H$_{2}$ outflow (MHO 2431A--D).  The white part in the 24~$\micron$ source is saturated.}
%contour levels: 8, 28, 48
\label{fig:G35.20-0.74}
\end{figure}

%1-25, f17+1
{\bf EGO G35.20$-$0.74} (Figure~\ref{fig:G35.20-0.74}).  EGO G35.20-0.74 is also known as G35.2N.  \citet{den84} reported that there are two compact \ion{H}{2} regions, located near the saturated 24~$\micron$ source.  In the figure, the lowest contour around the EGO roughly represents its shape.  Our continuum-subtracted H$_{2}$ image shows an hourglass outflow with a wide opening angle ($\sim40\arcdeg$) in the NE-SW direction.  In contrast to the extended H$_{2}$ emission, the 4.5~$\micron$ and $K$-band continuum emission are brighter at the base of the outflow and are similar in morphology.  The tips of the H$_{2}$ outflow are associated with 4.5~$\micron$ emission, but the 4.5~$\micron$ emission at the tips is much fainter than that near the base.  The peak of the $K$-band extended emission coincides with the 24~$\micron$ source.  %In addition, continuum emission is seen close to the base of the outflow.

\begin{figure}%18+1
\includegraphics[angle=0,scale=0.55]{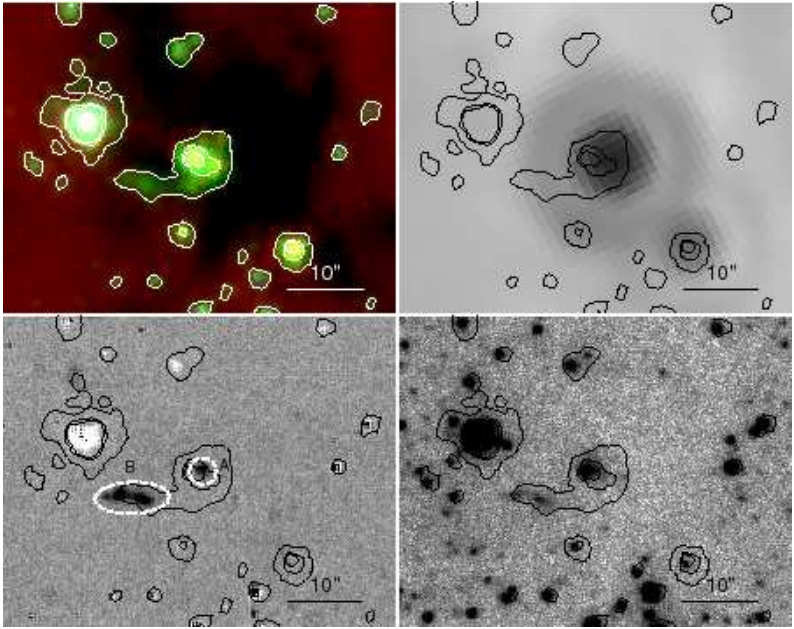}
\caption{Same as Figure~\ref{fig:G10.34-0.14}, but for EGO G35.68-0.18.  The dashed ellipse marks the lobe of the H$_{2}$ outflow (MHO 2432A--B).}
%contour levels: 5, 25, 45
\label{fig:G35.68-0.18}
\end{figure}

%1-26, f18+1
{\bf EGO G35.68$-$0.18} (Figure~\ref{fig:G35.68-0.18}).  A 24~$\micron$ source is located near the 4.5~$\micron$ peak of EGO G35.68-0.18.  The continuum-subtracted H$_{2}$ image shows two emission components: one close to the 4.5~$\micron$ peak (``A'' in the figure), the other extending toward the east (``B'').  The position of ``A'' is offset from the 4.5~$\micron$ peak by $\sim$2$\arcsec$.  The position of ``B'' approximately matches a diffuse component in the 4.5~$\micron$ image. The $K$-band image shows emission similar to the H$_{2}$, but the peak position of the component close to ``A'' matches the 4.5~$\micron$ peak.  This component is much brighter than ``B'' at 4.5~$\micron$ and $K$-band, while ``A'' and ``B'' are of similar brightness in H$_{2}$ emission.

%a lobe of an H$_{2}$ outflow to the east which are likely part of an outflow.  In contrast to the H$_{2}$ emission, both 4.5~$\micron$ and $K$-band image show intensive emission close to the base of the outflow.  The morphologies of 4.5~$\micron$, $K$-band, and H$_{2}$ emission are similar.
%There is an H$_{2}$ outflow located $\sim$30$\arcsec$ south of the EGO, but it is almost not detected in the $Spitzer$ image. %------\citet{lee11}.

\begin{figure}%19+1
\includegraphics[angle=0,scale=0.5]{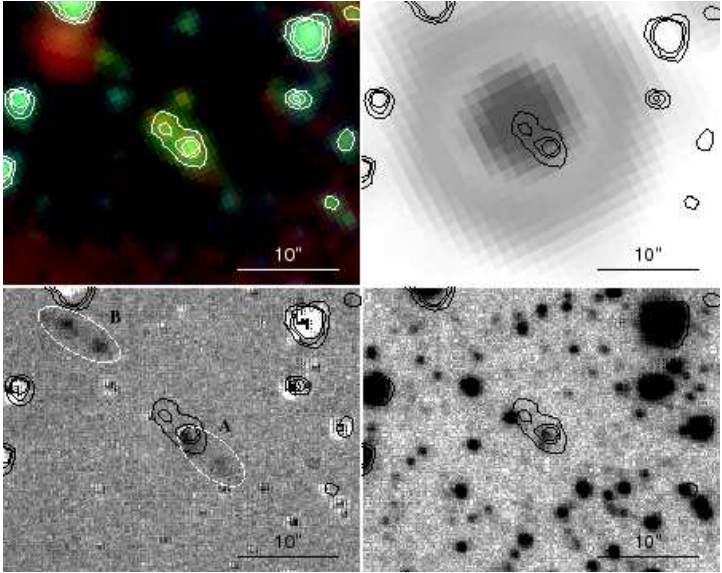}
\caption{Same as Figure~\ref{fig:G10.34-0.14}, but for EGO G35.79-0.17.  The dashed ellipses mark the H$_{2}$ outflow (MHO 2433A--B).}
%contour levels: 13, 23, 33
\label{fig:G35.79-0.17}
\end{figure}

%1-27, f19+1
{\bf EGO G35.79$-$0.17} (Figure~\ref{fig:G35.79-0.17}).  In the 4.5~$\micron$ contour map of EGO G35.79-0.17, there are two peaks with a separation of 3$\arcsec$.  The NE peak is associated with a 24~$\micron$ source.  The continuum-subtracted H$_{2}$ image exhibits an H$_{2}$ outflow in the NE-SW direction with the NE lobe brighter than the SW one.  In contrast to the more extended H$_{2}$ emission, the GLIMPSE image only shows bright emission close to the 24~$\micron$ source.  The $K$-band continuum image shows an extended source at the position of the SW 4.5~$\micron$ peak which does not coincide with the position of the 24~$\micron$ source.

%EGO G35.79-0.17 is located on an IRDC.  Figure~\ref{fig:G35.79-0.17} exhibits an H$_{2}$ outflow.  In contrast to more extended H$_{2}$ outflow, the 4.5~$\micron$ contour and $K$-band image show intensive emission close to the base of the outflow.  %The 4.5~$\micron$ emission on the base of the outflow is brighter that of the outflow.

\begin{figure}%20+1
\includegraphics[angle=0,scale=0.5]{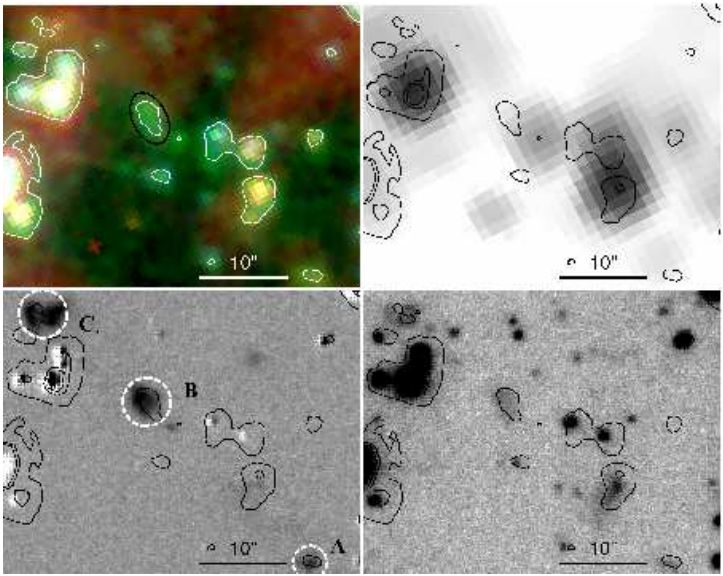}
\caption{Same as Figure~\ref{fig:G10.34-0.14}, but for EGO G35.83-0.20.  The black circle labels the position of the EGO.  The dashed circles mark the positions of the H$_{2}$ knots of the outflow in the NE-SW direction (MHO 2434A--C).}
%contour levels: 7, 37, 67
\label{fig:G35.83-0.20}
\end{figure}

%4-15, f20+1
{\bf EGO G35.83$-$0.20} (Figure~\ref{fig:G35.83-0.20}).  There are several faint 24~$\micron$ sources around EGO G35.83-0.20, but it is not clear which one is physically related to the EGO.  The continuum-subtracted H$_{2}$ image shows a chain of a few H$_{2}$ knots (MHO 2434A-C) in the NE-SW direction.  One of the H$_{2}$ knot (MHO 2434B) coincides with the EGO.  These H$_{2}$ knots are also seen in the $Spitzer$ image, however, the $Spitzer$ image contains a significantly larger number of emission features not seen in the continuum-subtracted H$_{2}$ image.  This makes the presence of the outflow ambiguous in the $Spitzer$ image.  In the $K$-band image, there is diffuse emission associated with the EGO.

\begin{figure}%21+1
\includegraphics[angle=0,scale=0.55]{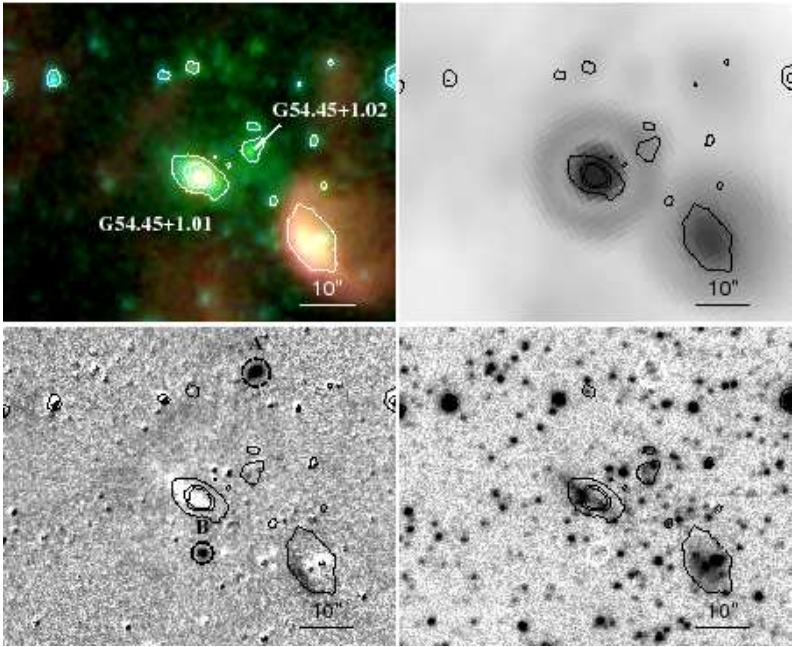}
\caption{Same as Figure~\ref{fig:G10.34-0.14}, but for EGO G54.45+1.01 and G54.45+1.02.  The positions of the two EGOs are labeled.  The dashed circles mark the positions of the H$_{2}$ knots (MHO 2622A--B).}
%contour levels: 10, 40, 70
\label{fig:G54.45+1.01}
\end{figure}

%3-14, f21+1
{\bf EGO G54.45$+$1.01 and G54.45+1.02} (Figure~\ref{fig:G54.45+1.01}).  A 24~$\micron$ source is associated with G54.45+1.01.  There are two H$_{2}$ knots near the EGO to the north west and south east located $\sim25\arcsec$ and $\sim10\arcsec$ away from the 24~$\micron$ source, respectively.  The alignment of these knots, and the lack of other nearby 24~$\micron$ sources, implies that they are associated with EGO G54.45+1.01.  A $K$-band continuum extended source is related to the EGO.  The orientation of the $K$-band extended source resembles that of the 4.5~$\micron$ emission.  For G54.45+1.02, an extended $K$-band source is associated with the 4.5~$\micron$ emission; however no 24~$\micron$ source is seen.

%EGO G54.45+1.02: 1-35

\begin{figure}%22+1
\includegraphics[angle=0,scale=0.55]{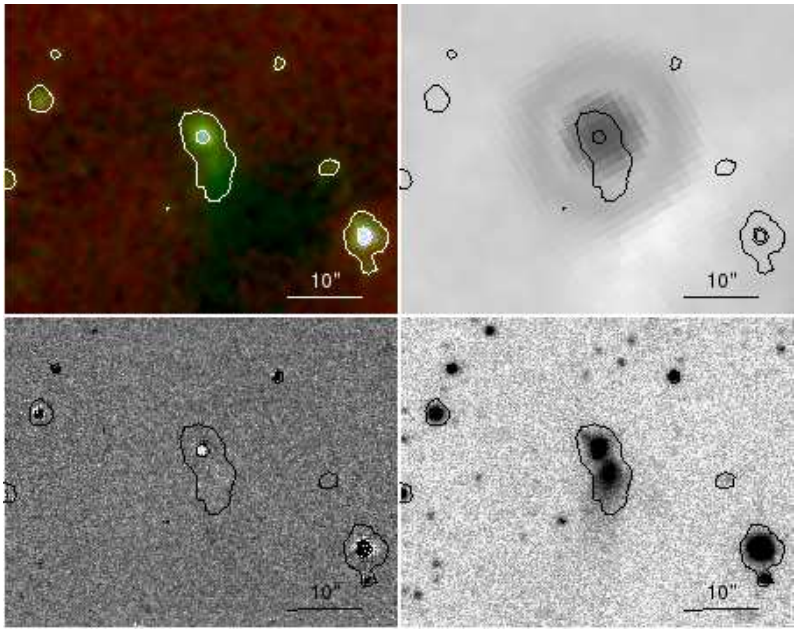}
\caption{Same as Figure~\ref{fig:G10.34-0.14}, but for EGO G58.09-0.34.}
%contour levels: 5, 35, 65
\label{fig:G58.09-0.34}
\end{figure}

%1-37, f22+1
{\bf EGO G58.09$-$0.34} (Figure~\ref{fig:G58.09-0.34}). EGO G58.09-0.34 is associated with a 24~$\micron$ source which coincides with the peak of the 4.5~$\micron$ emission.  The continuum-subtracted H$_{2}$ image does not show any H$_{2}$ features, but does show continuum emission with negative valued feature at the 4.5~$\micron$ peak.  In the $K$-band image, there are two extended sources with a separation of 3$\arcsec$ to the north and south.  The northern source is located on the 4.5~$\micron$ peak.  In the continuum-subtracted H$_{2}$ image, the northern source shows negative valued feature, but the southern one does not.

%\acknowledgments

%This research is based on observations with $AKARI$, a JAXA project with the participation of ESA.

\clearpage

\clearpage

%\clearpage

\clearpage

\clearpage

\clearpage

\clearpage

%\begin{figure}
%\includegraphics[angle=0,scale=0.8]{masers.eps}
%\caption{Methanol maser distributions around EGO G11.92-0.61 and G19.01-0.03.  The red dots mark the positions of methanol masers, and are superpose on the continuum-subtracted H$_{2}$ images.  The dashed ellipses label the H$_{2}$ outflows.}
%\label{fig:masers}
%\end{figure}

%\clearpage

%\clearpage
%\begin{figure}%
%\includegraphics[angle=0,scale=0.9]{hist.eps}
%\caption{Histogram of EGOs with and without H$_{2}$ detections.  }
%\label{fig:hist}
%\end{figure}

\begin{deluxetable}{lrcccl}
\tablecaption{H$_{2}$, $K$-, and $H$-band results}
\tablewidth{0pt}
\tabletypesize{\footnotesize}
\rotate
\tablehead{ & & \multicolumn{3}{c}{Detection} & \\
\cline{3-5}\noalign{\smallskip}
\colhead{EGO} & \colhead{D.} & \colhead{H$_{2}$} & \colhead{$K$-band} & \colhead{$H$-band} & \colhead{Note}\\
 & \colhead{(kpc)} & (Y/N) & (Y/N) & (Y/N) & }

\startdata
G10.29-0.13      & 2.2 & N & Y & N &                                                                \\%2-1, no BGPS detection
G10.34-0.14      & 2.0 & N & Y & N & continuum emission near the EGO                                \\%2-2, BGPS
G11.92-0.61      & 3.9 & Y & Y & Y & an H$_{2}$ outflow lobe                                        \\%1-1, outside of BGPS region
G12.02-0.21      &17.0 & Y?& N & N & an H$_{2}$ knot located at the north of the EGO                \\%1-2, BGPS
G12.20-0.03      & 4.6 & N & Y & N & continuum emission near the EGO                                \\%4-1, BGPS
G12.42+0.50      & 2.4 & Y?& Y & Y & extended H$_{2}$ emission, HII [UHP2009] G012.4180+00.5038     \\%4-2, BGPS
G12.68-0.18      & --  & N & N & N &                                                                \\%4-3, BGPS
G12.91-0.03$^{a}$& 4.7 & Y?& Y & N & a faint H$_{2}$ knot                                           \\%1-3, BGPS
G14.63-0.58$^{a}$& 2.2 & N & N & N &                                                                \\%1-5, BGPS
G16.58-0.08      & 3.6 & N & N & N &                                                                \\%3-2, BGPS
G16.59-0.05      & 4.4 & Y?& N & N & an H$_{2}$ knot                                                \\%2-3, BGPS
G16.61-0.24$^{a}$& --  & Y & Y & N & a faint H$_{2}$ outflow lobe                                   \\%1-6, no BGPS detection
G18.67+0.03      & 5.0 & N & Y & N &                                                                \\%1-7, no BGPS detection
G18.89-0.47$^{a}$& 4.5 & N & N & N &                                                                \\%1-8, BGPS
G19.01-0.03$^{a}$& 4.3 & Y & Y & N & an H$_{2}$ outflow with two H$_{2}$ knots                      \\%1-9, BGPS
G19.61-0.12      & 4.1 & N & Y & N & continuum emission near the EGO                                \\%2-5, no BGPS detection
G19.61-0.14      & --  & N & Y & N &                                                                \\%4-5, BGPS
G19.88-0.53$^{a}$& 3.4 & Y & Y & Y & a bipolar H$_{2}$ outflow, IRAS 18264-1152                     \\%1-10, BGPS
G20.24+0.07      & 4.6 & N & Y & N & continuum emission near the EGO                                \\%4-6, 
G28.83-0.25      & 5.1 & N & Y & Y & continuum emission near the EGO                                \\%1-19& --  &          \\% x
G28.85-0.23      & --  & N & N & N &                                                                \\%4-12, no BGPS detection
G35.04-0.47      & 3.4 & Y & N & N & a bipolar H$_{2}$ outflow                                      \\%&11.3 &          \\% x %1-22
G35.13-0.74$^{a}$& 2.4 & Y & Y & Y & a bipolar H$_{2}$ outflow                                      \\%& --  &          \\% x %1-23
G35.15+0.80$^{a}$& 4.7 & Y & Y & N & a set of aligned H$_{2}$knots                                  \\%, YSO is located at a giant pillar & --  & \\% x %1-24
G35.20-0.74$^{a}$& 2.4 & Y & Y & --& G35.2-0.74N, an hourglass shape bipolar H$_{2}$ outflow        \\%1-25 *** 2.3Myr, L=13.6 L
G35.68-0.18$^{a}$& 2.1 & Y & Y & Y & an H$_{2}$ outflow lobe                                        \\%1-26  ***age=8.7Myr, L=0.3 L
G35.79-0.17      & --  & Y & Y & N & a bipolar H$_{2}$ outflow                                      \\%& 7.9 &          \\% x %1-27
G35.83-0.20      & 2.1 & Y & Y & N & a set of aligned H$_{2}$ knots                                 \\%4-15
G36.01-0.20      & 5.9 & N & N & N &                                                                \\%1-28, BGPS
G54.11-0.04      & 4.9 & N & N & N &                                                                \\%4-23, no BGPS detection
G54.11-0.05      & --  & N & N & N &                                                                \\%4-24, BGPS
G54.45+1.01      & 3.7 & Y & Y & N & an H$_{2}$ outflow with two H$_{2}$ knots                      \\%& --  &          \\%  %3-14 age=4.7 Myr, L=1.4 L
G54.45+1.02$^{a}$& --  & N & Y & N &                                                                \\%& --  &          \\% x %1-35
G58.09-0.34$^{a}$& --  & N & Y & Y & continuum emission near the EGO                                \\%& --  &          \\% o %1-37
%34: 
%H2: 12Y+4Y?+18N
%K: 23Y
%H: 9Y

%G45.47+0.05 & --  &   & without K image\\% x %1-33
%9
\enddata
\tablenotetext{a}{Preselected targets before the H$_{2}$ observations}
\label{tab:h2}
\end{deluxetable}

\begin{deluxetable}{lrll}
\tablecaption{New MHO numbers in this paper}
\tablewidth{0pt}
%\rotate
\tablehead{\colhead{MHO number} & \colhead{RA (J2000)} & \colhead{DEC (J2000)} & \colhead{Note}}

\startdata

2303  & 18 13 57.5 & -18 54 29 & an H$_{2}$ outflow lobe from EGO G11.92-0.61      \\%1-1, Sgr

2243  & 18 21 53.6 & -14 35 50 & an H$_{2}$ outflow lobe from EGO G16.61-0.24      \\%1-6, Ser
2244  & 18 25 44.8 & -12 22 46 & a pair of H$_{2}$ knots near EGO G19.01-0.03      \\%1-9, Ser
2245  & 18 29 14.8 & -11 50 08 & a faint H$_{2}$ knot near EGO G19.88-0.53         \\%new in IRAS 18264, Ser

2429  & 18 56 58.1 & +01 39 37 & a bipolar outflow from EGO G35.04-0.47            \\%1-22, Aqu
2430  & 18 52 36.6 & +02 20 26 & a chain of H$_{2}$ knots from EGO G35.15+0.80     \\%1-24, Aqu
2431  & 18 58 12.9 & +01 40 33 & an hourglass H$_{2}$ outflow from EGO G35.20-0.74 \\%1-25, Aqu
2432  & 18 57 05.0 & +02 22 00 & an H$_{2}$ outflow lobe from EGO G35.68-0.18      \\%1-26, Aqu
2433  & 18 57 16.7 & +02 27 56 & a bipolar outflow from EGO G35.79-0.17            \\%1-27, Aqu
2434  & 18 57 26.9 & +02 29 00 & a chain of H$_{2}$ knots from EGO G35.83-0.20     \\%4-15, Aqu

2622  & 19 28 26.4 & +19 32 15 & two H$_{2}$ knots near EGO G54.45+1.01            \\%3-14, Vul

\enddata
\label{tab:mho}
\end{deluxetable}

\end{document}